\documentclass[
  prl,
  twocolumn,
  superscriptaddress,
  citeautoscript,
  showpacs,
  amsart,
  longbibliography
]{revtex4-2}

\usepackage{amsmath,bm,amsfonts}
\usepackage{physics,mathtools}
\usepackage{graphicx}
\usepackage[normalem]{ulem}
\usepackage{titlesec}
\usepackage[colorlinks=true,linkcolor=blue,citecolor=blue,urlcolor=blue]{hyperref}

\usepackage[dvipsnames]{xcolor}

\newcommand{\pdagger}{\phantom{\dagger}}

\usepackage{pdfpages}
\usepackage{pgffor}
\makeatletter
\AtBeginDocument{\let\LS@rot\@undefined}
\makeatother

\begin{document}

\title{Quench spectroscopy of amplitude modes in a one-dimensional critical phase}

\author{Hyunsoo Ha}
\affiliation{Department of Physics, Princeton University, Princeton, NJ 08544, USA}
\author{David A. Huse}
\affiliation{Department of Physics, Princeton University, Princeton, NJ 08544, USA}
\author{Rhine Samajdar}
\affiliation{Department of Physics, Princeton University, Princeton, NJ 08544, USA}
\affiliation{Department of Electrical and Computer Engineering, Princeton University, Princeton, NJ 08544, USA}

\begin{abstract}
We investigate the emergence of an amplitude (Higgs-like) mode in the gapless phase of the $(1+1)$D XXZ spin chain. Unlike conventional settings where amplitude modes arise from spontaneous symmetry breaking, here, we identify a symmetry-preserving underdamped excitation on top of a Luttinger-liquid ground state.  Using nonequilibrium quench spectroscopy, we demonstrate that this mode manifests as  oscillations of U(1)-symmetric observables following a sudden quench. By combining numerical simulations with Bethe-ansatz analyses, we trace its microscopic origin to specific families of string excitations. We further discuss experimental pathways to detect this mode in easy-plane quantum magnets and programmable quantum simulators. Our results showcase the utility of quantum quenches as a powerful tool to probe collective excitations, beyond the scope of linear response.
\end{abstract}

\maketitle

\emph{Introduction.}---When a continuous symmetry is spontaneously broken, the key collective excitations of the system are massless Goldstone modes and, in some cases, a massive amplitude (or Higgs-like) mode~\cite{pekker2015amplitude}. The latter corresponds to oscillations in the \textit{magnitude} of the order parameter, about its equilibrium value, without ``rotating'' it along the direction(s) of the continuous symmetry~\cite{higgs1964broken}. 

 In the solid state, such underdamped Higgs-like oscillations have been measured in a wide variety of systems ranging from superconductors~\cite{varma2002higgs,sooryakumar1980raman,littlewood1982amplitude,buzzi2021higgs} to quantum antiferromagnets~\cite{ruegg2008quantum,hong2017higgs}, and charge-density-wave-ordered compounds~\cite{pouget1991neutron,ren2004ultrafast,yusupov2010coherent}. In contrast to the 
Goldstone modes, whose decay rate vanishes faster than their frequency in the long-wavelength limit,  
the amplitude mode is in general damped via its decay into multiple lower-energy excitations. 
Therefore, the presence of an underdamped amplitude mode---especially in the vicinity of (gapless) quantum critical points (QCPs)---is an important question and has been extensively investigated for O($\mathcal{N}$) field theories, with $\mathcal{N}$\,$=$\,$2,3$, in $d$\,$=$\,$2$ and $3$ spatial dimensions~\cite{lindner2010conductivity,podolsky2011visibility,podolsky2012spectral,gazit2013fate}. 

Over the last decade, systems of ultracold bosonic and fermionic atoms  have offered a complementary platform to study the physics of such collective excitations~\cite{endres2012higgs,pollet2012higgs,leonard2017monitoring,behrle2018higgs,guo2019low,geier2021exciting}.
In particular, a recent experiment on a Rydberg atom array~\cite{Manovitz_2025,coarsening}, which observed long-lived oscillations of the order parameter, suggested that an emergent amplitude mode of the system could be excited via nonequilibrium quantum dynamics. Inspired by these technological developments and recent results, we ask: in what other gapless systems, 
in terms of $\mathcal{N}$ and $d$, 
can one observe an underdamped collective amplitude mode? 

\begin{figure}[t]
\centering
\includegraphics[width=0.95\linewidth]{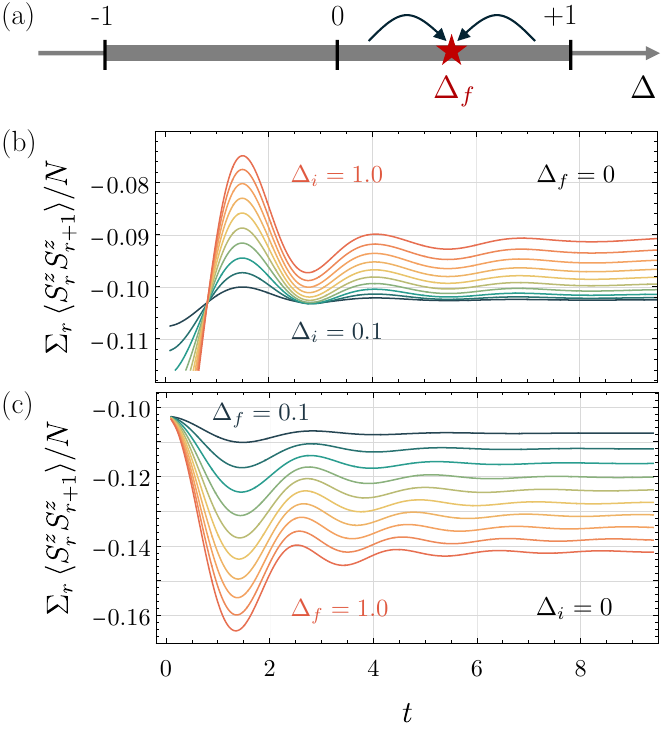}
\caption{(a) Quench protocol: the system is initialized in the ground state of the XXZ Hamiltonian ${H}(\Delta_i)$ and subsequently evolved under ${H}_f$\,$=$\,${H}(\Delta_f)$. (b) Dynamics of the U(1)-symmetric probe $\mathcal{O}^{zz}$ after a quench from $\Delta_i = 0.1,\ldots, 1.0$ (in steps of 0.1) to the free-fermion point $\Delta_f=0$. Similar underdamped oscillations are found for other $|\Delta_f|\le 1$ as well. (c) Dynamics after quenches from $\Delta_i = 0$ to $\Delta_f = 0.1,\ldots, 1.0$. The oscillation frequency is determined by $\Delta_f$, while its amplitude is set by $\lvert \Delta_f-\Delta_i \rvert$.}
\label{fig:protocol}
\end{figure}

To address this question, we begin by investigating the existence of a possible Higgs resonance at the QCP in a ($1+1$)D $\mathcal{N}$\,$=$\,$2$ system, namely, the paradigmatic XXZ spin chain (defined below). In the easy-plane regime of this model, the ground state is a gapless Luttinger liquid, characterized by power-law-decaying correlations~\cite{haldane1981luttinger}. The system's global U(1) symmetry, associated with spin rotations about the $z$-axis, remains unbroken in this critical phase. Importantly, however, an amplitude mode need not always be a direct consequence of spontaneous symmetry breaking. In general, any set of low-energy excitations can be classified according to how they transform under a given symmetry. An amplitude mode then simply represents fluctuations that preserve the symmetry of the underlying phase, i.e., its transformation properties are identical to those of the ground state.

With this definition of the amplitude mode as an underdamped symmetry-preserving resonance in mind, in this article, we uncover its existence in the XXZ model and demonstrate how it can be excited using quantum quenches that can drive the system far from equilibrium.
To probe these collective dynamics, we choose an observable that commutes with all the symmetries that the initial and final Hamiltonians share, and observe the time evolution of the expectation value of this probe operator after the quench.
Moreover, we theoretically identify the origin of this mode, both at the free-fermion point and in the interacting regime using the Bethe ansatz. Furthermore, in the Supplementary Material (SM), we show that the ($1+1$)D Ising model ($\mathcal{N}$\,$=$\,$1$), which has a $\mathbb{Z}_2$ symmetry instead of U(1), also hosts a similar collective excitation.  The \textit{experimentally accessible} quench spectroscopy methods that we develop here thus serve as a powerful tool---beyond linear response---in hunting for emergent collective phenomena.

\emph{Model and methods.}---We study an $N$-site spin-$1/2$ chain described by the antiferromagnetic  XXZ Hamiltonian
$
{H}(\Delta)=\sum_{r=1}^{N-1} J({S}^x_r{S}^x_{r+1}+{S}^y_r{S}^y_{r+1}+\Delta{S}^z_r{S}^z_{r+1})
$,
with exchange $J$\,$>$\,$0$. This model exhibits a U(1) symmetry since ${H}$ commutes with $\sum_r {S}^z_r$, and it is known to be gapless $\forall \,|\Delta|\le 1$. To understand whether there exists a collective mode with the same symmetry as the Hamiltonian, we focus on the simplest nonconserved U(1)-symmetric operator,
$
\mathcal{O}^{zz}=\sum_r {S}^z_r{S}^z_{r+1}/N,
\label{eqn:probe}
$
which is spatially uniform and therefore relevant for experiments in which one only has global control over the dynamics.

\begin{figure*}[tb]
    \includegraphics[width=0.9\linewidth]{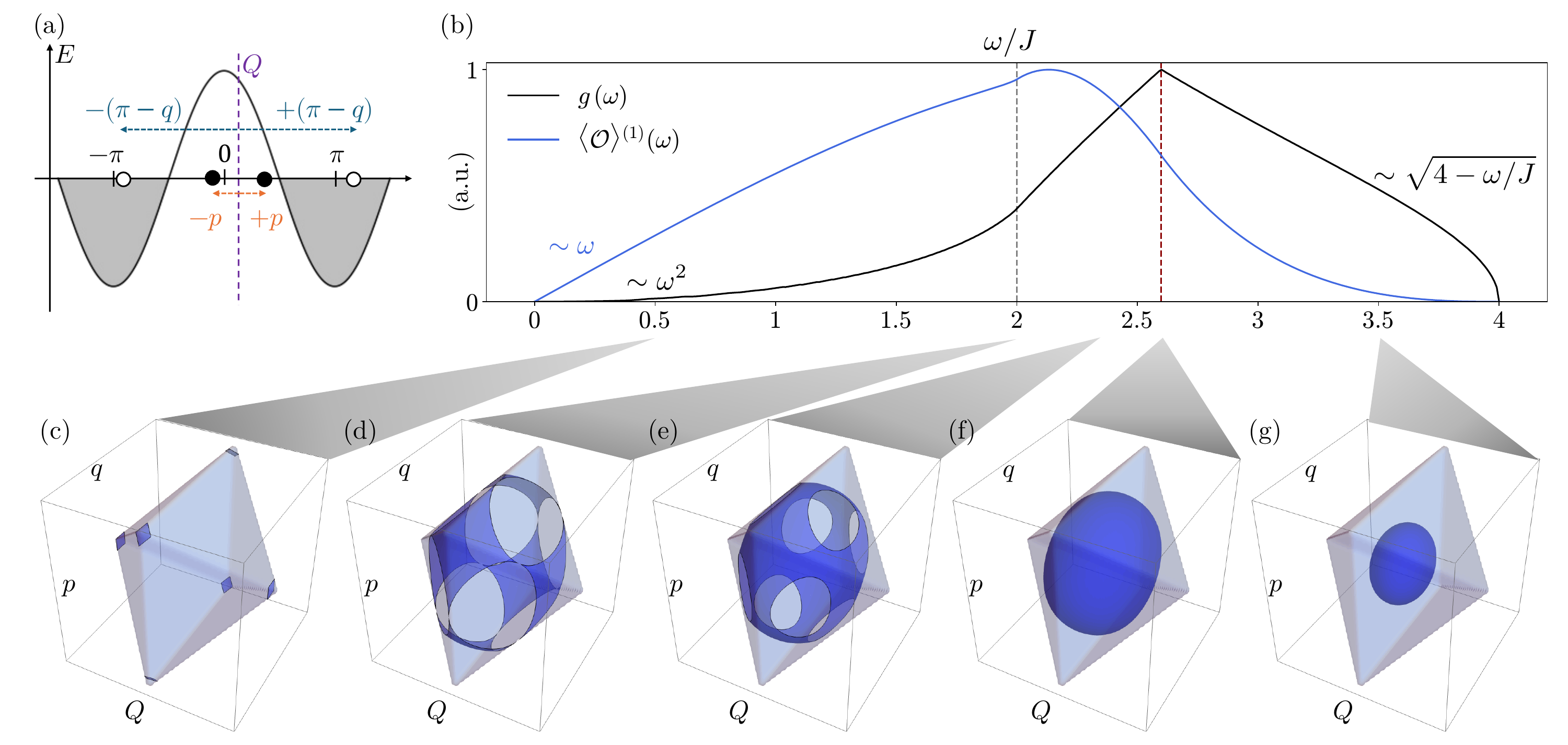}
    \caption{
    (a) Parametrization of the four crystal momenta $k_{p+}$, $k_{p-}$, $k_{h+}$, and $k_{h-}$ in terms of three parameters $Q$, $p$, and $q$, in consistency with momentum conservation. (b) Density of states $g(\omega)$ and Fourier component $\langle \mathcal{O}\rangle^{(1)}(\omega)$  as functions of $\omega/J$. (c--g) Surface morphology in the parameter space $(Q,p,q)$, where the energy difference $\varepsilon$ matches the frequency. {While the figure sketches $\mathcal{D}$, our calculations are restricted to $\mathcal{D}_{>0}$ to prevent double counting from the indistinguishability of $p\leftrightarrow-p$ and $q\leftrightarrow-q$}. Shown are selected values of $\omega/J$: (c) $0.5$, (d) $2$, (e) $2.3$, (f) $3\sqrt{3}/2\approx2.6$, and (g) $3.5$. The DOS is nondifferentiable at $\omega/J=2$ (gray dotted line) and exhibits a cusp at $\omega/J=3\sqrt{3}/2$ (red dotted line), corresponding to an extremal point where 
    {punctures} begin to appear on the surface as it attaches to the boundary of the domain $\mathcal{D}$ for $\omega/J < 3\sqrt{3}/2$.}
    \label{fig:dos}
\end{figure*}

Our protocol is as follows. We first prepare the ground state $\rvert\psi(0)\rangle$ of ${H}(\Delta_i)$. At $t$\,$=$\,$0$, we quench to ${H}_f$\,$\equiv$ \,${H}(\Delta_f)$ with $|\Delta_f|\le 1$, and subsequently evolve the state as $\rvert\psi(t)\rangle=\exp({-i{H}_f t})\rvert\psi(0)\rangle$. We then monitor the expectation value
$
\langle \mathcal{O}^{zz}\rangle_t=\langle \psi(t)|\mathcal{O}^{zz}\rvert\psi(t)\rangle .
$
Numerically, the initial state $\rvert\psi(0)\rangle$ is obtained using the density-matrix renormalization group (DMRG)~\cite{white1992density,white1993density,schollwock2005density}, while the real-time dynamics are simulated via the time-evolving block decimation (TEBD) algorithm~\cite{Vidal2003,Daley2004}, both with open boundary conditions. The good agreement of our results between system sizes $N=50$ and $N=100$, within the accessible simulation times, indicate that we are not limited by finite-size effects.

Interestingly, we observed pronounced, long-lived oscillations in $\langle \mathcal{O}^{zz}\rangle_t$, signaling the emergence of a coherent, underdamped collective mode. The oscillation frequency depends solely on $\Delta_f$, whereas the initial state determines the oscillation amplitude. At the free-fermion point $\Delta_f=0$ [Fig.~\ref{fig:protocol}(b)], the period of the dynamics is $T\approx 2.5/J$ (corresponding to an angular frequency $\omega=2\pi/T\approx2.5J$), essentially independent of $\Delta_i$. The amplitude thereof can be sizable: for instance, with $\Delta_i=1$ the oscillations reach nearly $20\%$ of the mean value. We emphasize that this oscillation is not simply due to the single-particle bandwidth~\cite{barmettler2010quantum} but rather, reflects a genuinely collective many-body mode. Similar oscillatory behavior persists across other values of $\Delta_f$, with the frequency drifting systematically as a function of $\Delta_f$ [Fig.~\ref{fig:protocol}(c)].

\emph{Quench spectroscopy.}---We now outline a general framework to interpret these oscillatory dynamics. Consider a local (but not necessarily single-site) operator $\mathcal{O}_r$ supported near site $r$. Its momentum-resolved form is defined as
$
    \mathcal{O}(p) =\sum_r e^{ipr}\mathcal{O}_r/N,
$
which selectively probes excitations with crystal momentum $p$. In what follows, we focus on the zero-momentum component $\mathcal{O}\equiv \mathcal{O}(p=0)$ that couples to excitations that share the same symmetry as the ground state, as for $\mathcal{O}^{zz}$ above. Translational invariance then enforces the selection rule
$
    \langle n|\mathcal{O}|m\rangle = \delta_{2\pi}(k_n - k_m)\langle n|\mathcal{O}|m\rangle ,
$
where $k_n$ and $k_m$ are the total momenta associated with eigenstates $|n\rangle$ and $|m\rangle$, defined via the action of the translation operator ${T}$ that shifts $r\to r+1$: 
$
{T}|n\rangle = e^{ik_n}|n\rangle$, ${T}|m\rangle = e^{ik_m}|m\rangle .
$
Here, $\delta_{2\pi}(k) \equiv\sum_{\ell\in\mathbb{Z}}\delta(k-2\pi \ell)$ is the Kronecker delta function defined up to periodic identification within the Brillouin zone.

We consider an initial state that retains significant overlap with the ground state  for $\Delta = \Delta_f$, $|0\rangle$~\cite{villa_Sanchez-Palencia_quenchspectroscopy,villa_Sanchez-Palencia_localquenchspectroscopy}. To leading order in $(\Delta_i-\Delta_f)$, the corresponding density matrix can be written as
$
    {\rho}\approx \rho^{}_{00}|0\rangle\langle0| + \sum_n (\rho^{}_{0n}|0\rangle\langle n| + \mathrm{h.c.} ) + \cdots ,
$
where the ellipsis denotes higher-order contributions. Under unitary evolution following a quench, the expectation value of $\mathcal{O}$ becomes
$
    \langle\mathcal{O}\rangle_t
    \approx \rho^{}_{00}\langle 0\lvert\mathcal{O}\rvert0\rangle  + \sum_{n}\delta^{}_{2\pi}(k_n)({\rho^{}_{0n}}e^{i(E_n-E_0)t}\langle n|\mathcal{O}|0\rangle + {\mathrm{c.c.}}) .
$
Taking the Fourier transform $\int_{-\infty}^{\infty} dt~e^{-i\omega t}\langle \mathcal{O}\rangle_t$ then yields
\begin{alignat}{1}
\label{eq:O(w)}
\langle \mathcal{O}\rangle(\omega) &\approx \sum_{{n}} \delta^{}_{2\pi}(k^{}_n)\,[\delta(\omega - E^{}_n + E^{}_0)\,\langle n|\mathcal{O}|0\rangle\,{\rho^{}_{0n}}\nonumber\\
&+\delta(\omega + E^{}_n - E^{}_0)\,\langle 0|\mathcal{O}|n\rangle\,{\rho^{}_{n0}}].
\end{alignat}
Generically, an underdamped mode is reflected as a singularity (often a pole or a branch cut) in the Fourier component $\langle \mathcal{O}\rangle(\omega)$ at some nonzero real frequency $\omega_\ast$, or at a complex frequency with a nonzero real part.
In the next section, we study excitations at fixed energy $\omega$ and compute the density of states 
$
g(\omega) \equiv \sum_{n} \delta_{2\pi}(k_n)\,\delta(\omega-(E_n-E_0))
$
at the free-fermion point $\Delta_f=0$. We then compare this with the Fourier component computed perturbatively for $\Delta_i\ll 1$, to characterize the oscillatory response observed in Fig.~\ref{fig:protocol}(b).

\emph{Noninteracting limit.}---For $\Delta=0$, the Hamiltonian reduces to
$
{H}=\sum_k E(k)c^\dagger_k c^{\pdagger}_k$, with $E(k)=J\cos k ,
$
via the Jordan-Wigner transformation. In this fermionic representation, the probe operator couples four momentum modes $\{k\}$:
\begin{equation*}
\mathcal{O}^{zz}
= \mathcal{C} + \frac{1}{N^2}\sum_{\{k\}}
\mathcal{A}(\{k\})\delta_{2\pi}(\kappa)\,
c_{k_{p+}}^\dagger c_{k_{p-}}^\dagger c^{\pdagger}_{k_{h+}} c^{\pdagger}_{k_{h-}} ,
\end{equation*}
where $\mathcal{C}$ collects constant terms and contributions proportional to the conserved charge $\sum_r S^z_r$, and
\begin{equation}
\mathcal{A}(\{k\})\equiv
\cos\!\left(\frac{\kappa}{2}\right)
\sin\!\left(\frac{k_{p+}-k_{p-}}{2}\right)
\sin\!\left(\frac{k_{h+}-k_{h-}}{2}\right).
\label{eqn:matrix_element_Ak}
\end{equation}
The quartic term creates two particles with crystal momenta $k_{p\pm}$ and annihilates two particles (i.e., creates two holes) with crystal momenta $k_{h\pm}$. The associated momentum transfer is $\kappa$\,$=$\,$(k_{p+}$\,$+$\,$k_{p-})$\,$-$\,$(k_{h+}$\,$+$\,$k_{h-})$, and momentum conservation is imposed modulo $2\pi$ by $\delta_{2\pi}(\kappa)$, allowing for Umklapp processes. The antisymmetry under exchange of identical fermions  is encoded in $\mathcal{A}(\{k\})$, which is odd under $k_{p+}\leftrightarrow k_{p-}$ and $k_{h+}\leftrightarrow k_{h-}$.

\begin{figure*}
    \centering
    \includegraphics[width=0.75\linewidth]{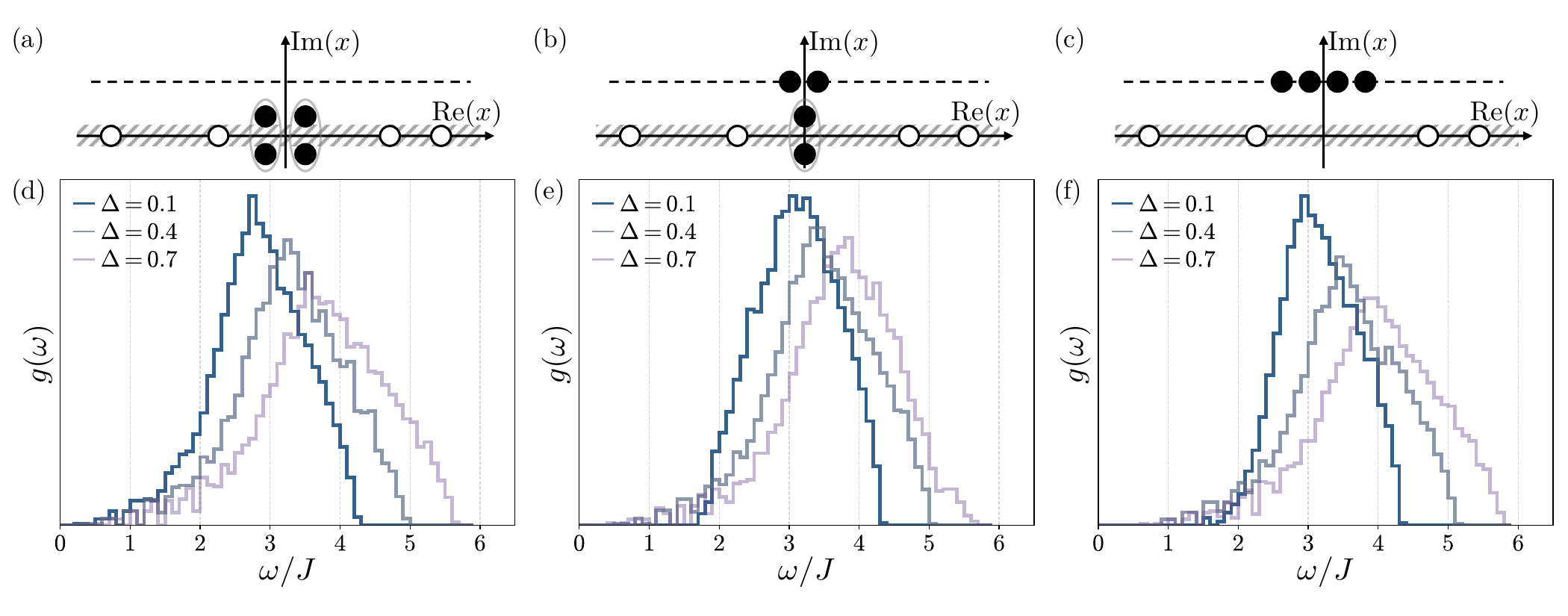}
    \caption{(a--c) Schematic illustrations of the three excitation types (I--III) discussed in the main text. Open circles denote real-axis holes, while filled circles represent 2-strings or length-1 complex rapidities. (d--f) Density of states $g(\omega)$ for $\Delta \in \{0.1, 0.4, 0.7\}$ for each excitation family. For small $\Delta$ (e.g., $\Delta=0.1$), $g(\omega$) for the type-I family smoothly connects to the free-fermion DOS: the peak coincides with the free-fermion value, and the spectrum extends continuously down to $\omega=0$.}
    \label{fig:Bethe_DOS}
\end{figure*}

To analyze the excitation spectrum, we parametrize the four momenta using three variables $Q$, $p$, and $q$ [Fig.~\ref{fig:dos}(a)]:
$
    k_{p\pm} = Q\pm p$, $k_{h\pm}=Q\pm(\pi-q) .
    \label{eqn:Qpq}
$
Their allowed domain $\mathcal{D}$ is defined by
$Q\in [-\pi/2,\pi/2]$,
$p \pm Q\in [-\pi/2,\pi/2]$, and
$Q\pm q \in [-\pi/2,\pi/2]$.
To avoid multiple counting due to indistinguishability between the two particles and between the two holes, we specifically consider the restricted domain $\mathcal{D}_{>0}=\{(Q,p,q)\,|\,p>0,q>0\}$.
The density of states for excitations at energy $\omega$ is then
$
    g(\omega) = \int_{\mathcal{D}_{>0}} dQ\,dp\,dq ~\delta(\omega-\varepsilon) ,
$
and the excitation energy is
$
\varepsilon 
= E(k_{p+})+E(k_{p-})-E(k_{h+})-E(k_{h-})
= 2J\cos Q\,(\cos p + \cos q) .
$
Thus, $g(\omega)$ is a surface integral of the Jacobian $|\nabla \varepsilon|^{-1}$ over the constant-energy surface defined by
$
\omega=\varepsilon=2J\cos Q\,(\cos p+\cos q)
\label{eq:surface_equation}
$
within $\mathcal{D}_{>0}$.
For comparison, we evaluate Eq.~\eqref{eq:O(w)} by computing $\rho_{0n}$ to first order in $\Delta\ll 1$, which gives 
$
    \langle \mathcal{O}\rangle^{(1)}(\omega) = \int_{\mathcal{D}_{>0}} dQ\,dp\,dq ~\delta(\omega-\varepsilon)\,|\mathcal{A}(Q, p, q)|^2/\varepsilon,
$
where the matrix element $\mathcal{A}(Q,p,q)=\sin p\,\sin q$ is obtained by substituting $Q$, $p$, and $q$ into Eq.~\eqref{eqn:matrix_element_Ak}
(see SM for details).

Figure~\ref{fig:dos}(b) plots the normalized DOS $g(\omega)$ (black) and $\langle \mathcal{O}\rangle^{(1)}(\omega)$ (blue) while Figs.~\ref{fig:dos}(c-g) show the corresponding constant-energy surfaces. As $\omega$ decreases from $4J$, the surface grows from the single point $(Q,p,q)=(0,0,0)$ into an expanding ellipsoid-like shape [Fig.~\ref{fig:dos}(g)] until it first touches the boundary of $\mathcal{D}$ [Fig.~\ref{fig:dos}(f)]. That contact produces a Van Hove singularity and a cusp in $g(\omega)$ at $\omega_{1*}/J=3\sqrt{3}/2\approx2.6$ (see SM). A second morphological change occurs at $\omega_{2*}/J=2$ [Fig.~\ref{fig:dos}(e)$\to$(c)], which again produces a Van Hove singularity. We also have singularities at the bottom ($\omega_{0*}=0$) and the top ($\omega_{4*}=4J$) of the band.

The structure of the constant-energy surface controls both the early-time behavior and the universal late-time decay. Since $\mathcal{A}(Q,p,q)$ vanishes on the boundary lines $p=0$ and $q=0$ where the cusp forms, the maximum of $\langle\mathcal{O}\rangle^{(1)}(\omega)$ is shifted to frequencies below $\omega_{1*}$. This produces a pronounced underdamped oscillation in the TEBD data at early times. At late times, Van Hove singularities dominate and the amplitude decays as a power law: the leading term is $\sim t^{-2}$ from $\omega_{0*}=0$, while the oscillatory component arises from the next-to-leading terms $\sim t^{-5/2}$ associated with the singularities at $\omega_{1*}$ and $\omega_{2*}$. A detailed investigation using the perturbative initial state is provided in the SM.

\emph{Bethe ansatz for $|\Delta_f|$\,$<$\,$1$.}---
The spin-$1/2$ XXZ chain is integrable and thus, exactly solvable using the Bethe ansatz~\cite{sutherland1985,levkovich2016bethe}. For even system sizes, the antiferromagnetic Hamiltonian can be mapped to its ferromagnetic counterpart by a staggered $\pi$ rotation about the $z$-axis:
$
  {H}(J_{\textsc{FM}}=-J,\,\Delta_{\textsc{FM}}=-\Delta)
  = {U}\,{H}(J,\Delta)\,{U}^\dagger 
$,
where ${U}=\prod_{\text{even } r}{\sigma}^z_r$ and ${\sigma}^z$ is the Pauli-$z$ operator. In this ferromagnetic frame, the fully polarized state with all spins up can be chosen to serve as the vacuum. Flipping $M$ spins relative to this ``vacuum'' generates $M$ quasi-momenta $k_1,\dots,k_M$, setting the total magnetization to $S^z=N/2 - M$. Our focus lies on the zero-magnetization sector, corresponding to $M=N/2$.

It is convenient to express the quasi-momenta in terms of \emph{rapidities} $x_j$, which are the Bethe ansatz's spectral parameters labeling each flipped spin. Rapidities reformulate the Bethe equations in a simpler form, with $k_j$ given explicitly as a function of $x_j$. In the easy-plane regime $|\Delta_{\textsc{FM}}|<1$, we define $\Gamma \equiv \cos^{-1}(-\Delta_{\textsc{FM}})$, {where $\Gamma\in[0,\pi]$,}  and parametrize $k_j$ as
$
\exp(ik_j) = -{\sinh \tfrac{\Gamma}{2}(x_j+i)}/{\sinh \tfrac{\Gamma}{2}(x_j-i)}
$. 
While the ground state is built entirely from real rapidities, excited states may include complex rapidities. One such class takes the form
$x$\,$=$\,$y$\,$+$\,$i p_0$, $y\in\mathbb{R}$, $p_0$\,$=$\,${\pi}/{\Gamma}$,
with a fixed imaginary part. Another important class consists of bound-state, or \emph{string}, solutions. A length-$n$ string comprises $n$ rapidities with a common real part,
$
x^{(n)}_j = \alpha + i (n+1-2j)$, $j=1,\dots,n ,
$
centered at $\alpha\in\mathbb{R}$. In addition, $p_0$-shifted strings can be formed by adding $i p_0$. The admissible species of strings and their possible lengths depend on $\Delta$ (or equivalently, $\Gamma$).

To construct an eigenstate with $M$ rapidities, one specifies an ansatz distributing rapidities among the allowed species, and then solves the coupled Bethe equations under periodic boundary conditions~\cite{Takahashi_1999}. Each rapidity $x_j$ is associated with a Bethe quantum number (BQN) $I_j$, defined within its species. If $m$ rapidities belong to a given species, their BQNs are distinct integers when $m$ is odd and distinct half-odd integers when $m$ is even. A consistent solution of the $M$ equations yields a valid eigenstate. For example, the ground state corresponds to $M=N/2$ real rapidities with consecutive BQNs: $-(M-1)/2,\dots,(M-1)/2$. {Excited states can be generated by removing some of these quantum numbers from the filled real-rapidity sea. We refer to each missing real-rapidity BQN as a \textit{hole}. At $\Delta_{\mathrm{FM}}=0$, this coincides with the usual fermionic hole in the free-fermion description.} Additional technical details of the Bethe ansatz equation are described in the SM.

In order to enable a direct comparison with the TEBD numerics for $\Delta\geq 0$, we work in the ferromagnetic frame where $\Delta_{\textsc{FM}}\le 0$. At the free-fermion point $\Delta_{\textsc{FM}}=0$, string solutions are not permitted. In this case, the two-particle-two-hole excitation reduces to a configuration consisting of two complex rapidities together with two holes in the real-rapidity line. However, this configuration does not persist into the regime $\Delta_{\textsc{FM}}<0$~\cite{Takahashi_Suzuki1972,Takahashi_1999}, since complex rapidities do not carry free BQNs~\footnote{Equivalently, there is no spin-wave branch for $\Delta_{\textsc{FM}}<0$.}. More precisely, a configuration with $m$ complex rapidities is associated with the \textit{fixed} set of BQNs $I_c\in\{-(m-1)/2,\dots,(m-1)/2\}$.  
For the zero-momentum excitations of interest to us, the Bethe quantum numbers of the 2-strings are similarly pinned near zero. Specifically, we have $I_s=0$ for a single 2-string, while for two 2-strings, $I_s\in\{-1/2,1/2\}$~\footnote{These values can shift slightly at certain anisotropies, but we focus on the family that remains robust across the range considered.}. 

Guided by these considerations and the free-fermion analysis, we thus focus our attention on excitations characterized by four independent momenta, subject to the global constraint of momentum conservation. Configurations with fewer degrees of freedom fail to reproduce the oscillation frequency observed numerically (see SM). Based on these criteria, we identify three distinct zero-momentum excitation families [see Fig.~\ref{fig:Bethe_DOS}(a--c)]: (I) two 2-strings; (II) one 2-string combined with two complex rapidities; and (III) four complex rapidities with no strings. In all three families, the excitation relative to the ground state corresponds to removing four BQNs from the real-rapidity line (i.e., creating four holes),
with momentum conservation requiring that the removed BQNs sum to zero.

Figure~\ref{fig:Bethe_DOS} presents the DOS $g(\omega)$ as a function of energy for each excitation family, computed by solving the Bethe equations for $N$\,$=$\,$100$, matching the TEBD system size. 
All three excitation families show a pronounced DOS peak. However, only type-I (two 2-strings) produces a peak that matches the TEBD oscillation frequency as $\Delta_{\textsc{FM}}\to 0$. Additionally, type-I excitations also extend to zero energy, whereas types II and III involve complex rapidities with a finite energy cost, which opens a gap and shifts their resonance away from the TEBD value. Numerically, 2-strings and complex rapidities with BQNs at or very near zero ($\{0\}$ or $\{-0.5,,0.5\}$) become nearly costless in this limit; other choices retain a finite cost. Consistently, only the type-I family develops a zero-energy DOS tail, unlike types II and III, in line with the free-fermion result in Fig.~\ref{fig:dos}.

While this analysis identifies the excitation species responsible for the amplitude mode response, computing its frequency as a function of the anisotropy $\Delta$ remains an open question, owing to the fact that the Bethe ansatz does not allow for straightforward evaluation of matrix elements. As a result, one only has analytical access to $g(\omega)$ rather than $\langle\mathcal{O}\rangle(\omega)$, and this precludes a quantitative calculation of the resonance frequency (as we learned earlier from the nontrivial shift in Fig.~\ref{fig:dos}(b) in the free-fermion case).

\textit{Discussion and outlook.}---In this work, we reveal a novel 
amplitude mode in the gapless phase of the $(1$\,$+$\,$1)$D XXZ model and demonstrate how it can be accessed through quench spectroscopy. Quantum quenches in the XXZ model have been investigated extensively, particularly in the context of equilibration (or lack thereof) to the generalized Gibbs ensemble~\cite{caux2013time,wouters2014quenching,pozsgay2014correlations,goldstein2014failure,pozsgay2014failure}. However, most numerical studies have concentrated on the temporal decay of correlations in the antiferromagnetic phase~\cite{liu2014quench,ramos2023power} or on dynamics initiated from magnetic domain wall states~\cite{lancaster2010quantum,gruber2019magnetization}. In contrast, in the critical phase of the XXZ model, the ground state cannot be described by a small number of spin flips above a magnetically ordered state~\cite{brockmann2014neel,brockmann2014gaudin}. The amplitude mode we identify here thus appears to have previously gone unnoticed. We further clarify its microscopic origin by analyzing the Bethe ansatz and uncovering its connection to string excitations. An interesting direction for future work is to understand if a more refined understanding of these quench dynamics can be obtained, for instance, in terms of quasiparticle fractionalization~\cite{foster2011quantum} or bosonization~\cite{pollmann2013linear}.

Experimentally, the amplitude mode described herein can be accessed in several ways. Recent neutron scattering experiments have observed long-lived, quasiperiodically oscillating coherent quantum dynamics following a local quench in the Heisenberg antiferromagnet KCuF$_3$~\cite{scheie2022quantum}. These studies could naturally be extended to XXZ magnets in the easy-plane regime, such as Cs$_2$CoCl$_4$~\cite{breunig2013spin,breunig2015low,Laurell_2021}, which is well described by $\Delta$\,$\approx$\,$0.12$. More recently, superconducting qubit arrays~\cite{maruyoshi2023conserved,PhysRevLett.133.240403,rosenberg2024dynamics} and ultracold molecules in optical tweezers~\cite{anderegg2019optical,kaufman2021quantum, holland2023demand,cornish2024quantum} have emerged as highly tunable platforms for digital and analog quantum simulation of the XXZ model, respectively. The local addressability intrinsic to these systems enables new lines of inquiry. For example, our analysis identified several types of Bethe ansatz excitations with well-defined peaks in their DOS. We can then ask the question of which probes, beyond $\mathcal{O}^{zz}$, can couple to and see the responses associated with these different families, and whether these operators can be programmably engineered in such synthetic systems.

\begin{acknowledgments}
\textit{Acknowledgments.}---We thank Meigan Aronson, Leon Balents, Pradip Kattel, and Bella Lake for useful discussions. This research was supported in part by NSF QLCI grant OMA-2120757. R.S. was supported by the Princeton Quantum Initiative Fellowship. This work was performed in part at the Aspen Center for Physics, which is supported by a grant from the Simons Foundation (1161654, Troyer). This work was also supported in part by the Heising-Simons Foundation, the Simons Foundation, and National Science Foundation Grants No. NSF PHY-1748958 and PHY-2309135 to the Kavli Institute for Theoretical Physics (KITP).The TEBD calculations presented in this article were implemented using the \textsc{ITensor} library~\cite{itensor} and performed on  computational resources managed and supported by Princeton Research Computing, a consortium of groups including the Princeton Institute for Computational Science and Engineering (PICSciE) and the Office of Information Technology's High Performance Computing Center and Visualization Laboratory at Princeton University. 
\end{acknowledgments}

\bibliography{main}

\newpage


\section*{Supplementary material}

# Supplemental Material for "Quench spectroscopy of amplitude modes in a one-dimensional critical phase"

Hyunsoo Ha,[1] David A. Huse,[1] and Rhine Samajdar[1,2]

[1]*Department of Physics, Princeton University, Princeton, NJ 08544, USA*

[2]*Department of Electrical and Computer Engineering,*
*Princeton University, Princeton, NJ 08544, USA*

In this Supplemental Material, we first present a detailed analysis of the XXZ model at the free-fermion point, $\Delta = 0$. In particular, we perturbatively compute the spectral function, analyze the Van Hove singularities, and discuss how they shape the late-time dynamics. We then show that no Higgs mode appears at the $(1+1)$-dimensional Ising quantum critical point within the framework of linear-response theory. However, the out-of-equilibrium quench spectroscopy with selective symmetric probes does exhibit oscillations analogous to those in the XXZ model. Finally, we describe the excitations of the easy-axis XXZ model with $|\Delta| < 1$ through the Bethe ansatz formulation.

## SI. XXZ MODEL AT THE FREE-FERMION POINT

### A. Quench signal $\langle \mathcal{O} \rangle(\omega)$ for $\Delta \ll 1$

We prepare the initial state as the ground state $|\psi_\Delta\rangle$ of the XXZ Hamiltonian at anisotropy $\Delta$. For small $\Delta$, we treat the $S^z S^z$ term as a perturbation to the noninteracting XX point. Setting $J = 1$ for simplicity,

$$H(\Delta) = \sum_{r=1}^{N-1} \left( S_r^x S_{r+1}^x + S_r^y S_{r+1}^y + \Delta\, S_r^z S_{r+1}^z \right) = H_0 + \Delta\, N\, \mathcal{O}^{zz}, \tag{S1}$$

where $H_0$ is the XX Hamiltonian $H(\Delta = 0)$, and $\mathcal{O}^{zz} \equiv \sum_r S_r^z S_{r+1}^z / N$ as we defined in the main text, which is the probe operator being quenched. Let $\{|n\rangle, E_n\}$ be the eigenbasis of $H_0$ ordered by energy, with ground state $|0\rangle$ and $E_0$. To first order in $\Delta$,

$$|\psi_\Delta\rangle = \frac{1}{Z_\Delta} \left[ |0\rangle + \Delta\, N \sum_{n \neq 0} |n\rangle\, \frac{\langle n | \mathcal{O}^{zz} | 0 \rangle}{E_0 - E_n} + \mathcal{O}(\Delta^2) \right], \tag{S2}$$

where $Z_\Delta$ is the normalization factor.

The off-diagonal weight between $|0\rangle$ and an excited state $|n\rangle$ that participates in the spectroscopic response is

$$\rho_{0n} = \langle 0|\psi_\Delta\rangle \langle \psi_\Delta|n\rangle = \frac{\Delta N}{Z_\Delta^2} \frac{(\langle n | \mathcal{O}^{zz} | 0 \rangle)^*}{E_0 - E_n} + \mathcal{O}(\Delta^2). \tag{S3}$$

For $\omega > 0$, we derived in the main text that the Fourier component of the signal picked up by $\mathcal{O}^{zz}$ is simply

$$\langle \mathcal{O}^{zz} \rangle(\omega) \approx \sum_{n > 0} \delta_{2\pi}(k_n)\, \delta(\omega - E_n + E_0)\, \langle n | \mathcal{O}^{zz} | 0 \rangle\, \rho_{0n}. \tag{S4}$$

Inserting (S3) into this equation gives

$$\langle \mathcal{O}^{zz} \rangle(\omega) \approx \frac{\Delta N}{Z_\Delta^2} \sum_{n > 0} \delta_{2\pi}(k_n)\, \delta(\omega - E_n + E_0)\, \frac{|\langle n | \mathcal{O}^{zz} | 0 \rangle|^2}{E_0 - E_n} + \mathcal{O}(\Delta^2). \tag{S5}$$

Following the main text, we parametrize the relevant two-particle excited states by $(Q, p, q)$ and write the excitation energy as $\varepsilon(Q, p, q)$ and the matrix element as $\mathcal{A}(Q, p, q) = N^2 \langle n | \mathcal{O}^{zz} | 0 \rangle$. Going to the continuum and restricting ourselves to the positive-energy half of the domain $\mathcal{D}_{>0}$ to avoid double-counting, we have

$$\langle \mathcal{O}^{zz} \rangle(\omega) \approx \frac{\Delta}{Z_\Delta^2} \int_{\mathcal{D}_{>0}} dQ\, dp\, dq\, \delta(\omega - \varepsilon(Q, p, q))\, \frac{|\mathcal{A}(Q, p, q)|^2}{\varepsilon(Q, p, q)} + \mathcal{O}(\Delta^2). \tag{S6}$$

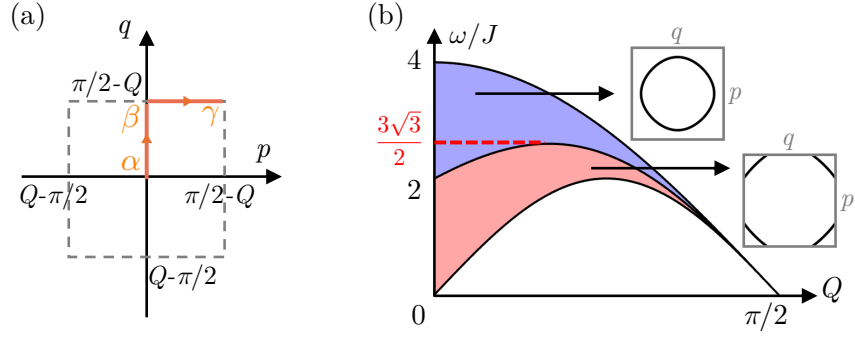

FIG. S1. (a) Range of $p$ and $q$ allowed for a fixed value of $Q$: $p, q \in [Q - \pi/2, \pi/2 - Q]$. (b) Illustration of Eq. (S9) (blue) and Eq. (S10) (pink). These curves indicate where the contour described in Eq. (S8) is connected or disconnected, given the restricted domain.

In the main text, we also define the first-order quantity

$$\langle \mathcal{O} \rangle^{(1)}(\omega) \equiv \int_{\mathcal{D}_{>0}} dQ \, dp \, dq \, \delta(\omega - \varepsilon(Q, p, q)) \, \frac{|\mathcal{A}(Q, p, q)|^2}{\varepsilon(Q, p, q)}, \tag{S7}$$

which is shown in Fig. 2 together with $g(\omega)$.

We note that for small $\omega \ll 1$, the density of states scales as $g(\omega) \sim \omega^2$. Thus, the integrand in (S7) behaves like $g(\omega)/\omega \sim \omega$, which does not diverge at the origin. Consequently, $\langle \mathcal{O} \rangle^{(1)}(\omega) \to 0$ as $\omega \to 0$ and exhibits a well-defined peak at finite $\omega$.

## B. Identifying the singularities

As discussed in the main text, the evolution of the constant-energy surface of excitations gives rise to Van Hove singularities. One of these produces the pronounced peak in $g(\omega)$ and strongly affects the Fourier component $\langle \mathcal{O} \rangle(\omega)$. Here, we locate the singular frequencies by characterizing the surface morphology in the parameter space $(Q, p, q) \in \mathcal{D}$. These reparametrized momenta $Q$, $p$, and $q$ satisfy

$$\omega = \varepsilon = 2J \cos Q \, (\cos p + \cos q), \tag{S8}$$

as described earlier. For fixed $Q$, the remaining parameters $p$ and $q$ lie in a square of side length $\pi/2 - |Q|$. Within this region, the energy $\varepsilon = 2J \cos Q (\cos p + \cos q)$ decreases monotonically along the orange path in Fig. S1(a), following the sequence $\alpha \to \beta \to \gamma$.

To analyze the contour defined by Eq. (S8), first consider its intersection with the line $\overline{\alpha\beta}$. For

$$2(1 + \sin Q) \cos Q \leq \omega/J \leq 4 \cos Q, \tag{S9}$$

the contour is a closed loop. This range is shown in blue in Fig. S1(b). When the contour intersects the line $\overline{\beta\gamma}$, the solution consists of four disconnected segments. This occurs for

$$4 \sin Q \cos Q \leq \omega/J \leq 2(1 + \sin Q) \cos Q, \tag{S10}$$

plotted in pink in Fig. S1(b). Outside these two regimes, there are no solutions.

The schematic in Fig. S1(b) illustrates the morphology of the constant-energy surface by drawing a horizontal line at a given $\omega/J$ and identifying which regions it intersects. The first singularity $\omega_F$ appears when the constant-energy surface meets the face of $\mathcal{D}$. This is the point where a hole opens: the horizontal line at fixed $\omega/J$ becomes a tangent to the common boundary between the pink and blue regions, which occurs at the maximum of $2(1 + \sin Q) \cos Q$. At $\sin Q = 1/2$,

$$\omega_F/J = \frac{3\sqrt{3}}{2} \approx 2.60, \tag{S11}$$

which produces the cusp in the density of states. The second singularity $\omega_E$ occurs when the constant-energy surface meets an edge of $\mathcal{D}$. Tangency to the lower boundary of the pink region gives $Q = \pi/4$, $|p| = |q| = \pi/4$, and

$$\omega_E / J = 2. \tag{S12}$$

There are also trivial singularities at the band top and bottom, where the surface collapses to a vertex of $\mathcal{D}$:

$$\omega_T = 4J, \qquad \omega_B = 0. \tag{S13}$$

Now, we investigate the typical value of the matrix element at the singularities, where the constant-energy surface touches the boundaries of the domain $\mathcal{D}$. In the XX model, which is the Jordan-Wigner-transformed free-fermion case, two particles and two holes are indistinguishable within each pair, so the matrix element is symmetric between pairs, $\mathcal{A}(Q, p, q) = \sin p \sin q$, where $2p$ and $2q$ are the inter-particle and inter-hole momentum differences. With $J = 1$, the dispersion is $\varepsilon(Q, p, q) = 2 \cos Q (\cos p + \cos q)$. Let $\delta \omega$ be the detuning from a singular frequency. The leading scalings near the singular points in the allowed domain $\mathcal{D}_{>0}$ are:

- Near $\omega_B$ (bottom of the band): The constant-energy surface consists of two types of separate regions, and both yield the same scaling of the matrix element. These are, first, $Q \approx 0$ with $p, q \approx \frac{\pi}{2} - Q$; and second, $Q \approx \frac{\pi}{2}$ with $p, q \approx 0$, with

$$\delta \omega \approx \delta Q \left( 2 - \frac{\delta p^2}{2} - \frac{\delta q^2}{2} \right). \tag{S14}$$

  A typical scaling is $\delta Q \sim \delta \omega$ and $\delta p \sim \delta q \sim \mathcal{O}(1)$. In both cases,

$$\mathcal{A}(Q, p, q) = \mathcal{O}(1). \tag{S15}$$

- Near $\omega_E$ (edge touch): Here $Q \approx \frac{\pi}{4}$ and $p, q \approx \frac{\pi}{2} - Q \approx \frac{\pi}{4}$, so

$$\mathcal{A}(Q, p, q) = \mathcal{O}(1). \tag{S16}$$

- Near $\omega_F$ (face touch): In the domain $\mathcal{D}_{>0}$, take $Q \approx \frac{\pi}{6}$, $p \approx \frac{\pi}{2} - Q \approx \frac{\pi}{3}$, and $q \approx 0$. Expanding

$$\omega_F + \delta \omega = 2 \cos \left( \frac{\pi}{6} + \delta Q \right) \left[ \cos \left( \frac{\pi}{3} - \delta Q + \delta p \right) + \cos (\delta q) \right] \tag{S17}$$

  gives the typical scales $\delta Q \sim \delta \omega$ and $|\delta q| \sim |\delta \omega|^{1/2}$. Hence,

$$\mathcal{A}(Q, p, q) \sim \sin q \sim |\delta \omega|^{1/2}. \tag{S18}$$

- Near $\omega_T$ (top of the band): All $Q, p, q$ are small. Then

$$|\delta \omega| \sim \mathcal{O}(\delta Q^2 + \delta p^2 + \delta q^2), \tag{S19}$$

  so $\delta Q, \delta p, \delta q = \mathcal{O}(|\delta \omega|^{1/2})$, and

$$\mathcal{A}(Q, p, q) \sim \sin p \sin q \sim |\delta \omega|. \tag{S20}$$

### C. Van Hove singularities in late-time dynamics

As mentioned in the main text, early-time oscillations are set by the peak of the Fourier component $\langle \mathcal{O} \rangle(\omega)$. A large DOS typically induces a proximate peak in $\langle \mathcal{O} \rangle(\omega)$, although they need not coincide because $\langle \mathcal{O} \rangle(\omega)$ is weighted by matrix elements. In the previous section, we located the frequencies where Van Hove singularities occur. Here, we show how these singularities set the late-time power-law envelope, with exponents fixed by the singularity type. Without such singularities, one would expect exponentially damped oscillations with a decay rate set by the width of the Fourier component; instead, the dynamics display algebraic decay controlled by the singularities.

Using $\langle \mathcal{O} \rangle^{(1)}(\omega)$ from Eq. (S7), we compute the time evolution

$$\langle \mathcal{O} \rangle_t = \int_{\omega_B = 0}^{\omega_T = 4J} d\omega \, \cos(\omega t) \, \langle \mathcal{O} \rangle^{(1)}(\omega), \tag{S21}$$

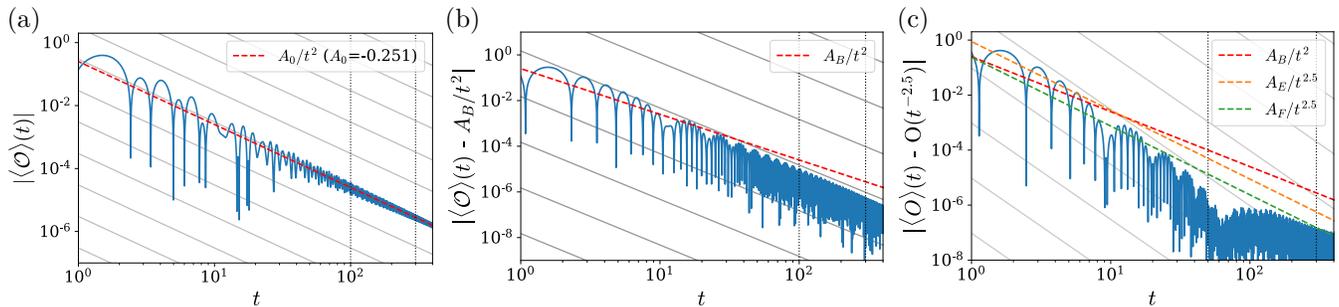

FIG. S2. (a) Log-log plot of $\langle\mathcal{O}\rangle_t$ for $t \in [1, 400]$. The envelope follows a power law $\sim t^{-2}$, parallel to the gray guide lines. (b) Residual after fitting $\langle\mathcal{O}\rangle_t \approx A_B t^{-2}$, shown on log-log axes; the red dotted curve displays $A_B t^{-2}$ and the fit window is $t \in [100, 300]$. (c) Residual after subtracting the $t^{-2}$ and $t^{-2.5}$ terms, shown on log-log axes for $t \in [50, 300]$. Reference curves $A_B t^{-2}$, $A_E t^{-2.5}$, and $A_F t^{-2.5}$ [see Eq. (S24)] are overlaid for comparison.

since $\langle\mathcal{O}\rangle^{(1)}(\omega)$ is real and the cosine arises from symmetric contributions at $\pm\omega$.

The leading contribution stems from $\omega_B$. Here, $g(\omega) \sim \omega^2$ and the matrix element is $\mathcal{O}(1)$, so

$$\langle\mathcal{O}\rangle^{(1)}(\omega) \sim g(\omega) \frac{|\mathcal{A}|^2}{\omega} \sim \omega, \tag{S22}$$

which yields an algebraic decay with exponent 2. In Fig. S2(a), the log-log plot follows a line of slope $-2$. Thus,

$$\langle\mathcal{O}\rangle_t \approx A_B t^{-2} + \mathcal{O}(t^{-2}). \tag{S23}$$

Figure S2(b) shows $A_B t^{-2}$ (red dotted) and the residual $|\langle\mathcal{O}\rangle_t - A_B t^{-2}|$ on log-log axes. In the fitting window $t \in [100, 300]$ the residual lies below the fit and itself follows a power law $t^{-2.5}$, as indicated by the gray guide lines. On a linear scale (not shown), the residual exhibits beating with period $\sim 11/J$, set by the two singular frequencies $\omega_E = 2J$ and $\omega_F = (3\sqrt{3}/2)J$, with a nearly constant relative amplitude over time, consistent with a common power-law envelope. Therefore,

$$\langle\mathcal{O}\rangle_t \approx A_B t^{-2} + t^{-2.5}[A_E \cos(\omega_E t + \phi_E) + A_F \cos(\omega_F t + \phi_F)] + \mathcal{O}(t^{-2.5}). \tag{S24}$$

In Fig. S2(c), the reference curves $A_B t^{-2}$, $A_E t^{-2.5}$, and $A_F t^{-2.5}$ are shown together with the residual magnitude. The remaining discrepancy decays faster, as $t^{-3.5}$, and is likewise captured by subleading singular contributions.

## SII.  Linear response theory of the (1+1)D Ising critical point

Motivated by recent experimental observations [1] studying $\mathcal{N} = 1$ systems, we investigate whether a Higgs resonance (i.e., an amplitude mode) can exist at the Ising quantum critical point (QCP) in (1 + 1)D. Specifically, we compute the dynamical structure factor of the $\sigma^x$ operator in the one-dimensional quantum transverse-field Ising model (TFIM), which possesses a $\mathbb{Z}_2$ symmetry generated by $\prod_\ell \sigma_\ell^x$. By virtue of universality, the presence of a Higgs resonance at the Ising QCP—if it exists—should be detectable within this exactly solvable model.

### A.  The transverse-field Ising model

We follow the approach of Dziarmaga [2] in this section. The (1 + 1)D TFIM is described by the Hamiltonian

$$H = -\sum_{\ell=1}^{N} \left(\sigma_\ell^z \sigma_{\ell+1}^z + g\sigma_\ell^x\right), \tag{S25}$$

for a chain of length $N$ with periodic boundary conditions $\sigma_{N+1} = \sigma_1$. For convenience, we take $N$ to be even. The system exhibits a quantum critical point at $g_c = 1$, which belongs to the Ising universality class. For $g < g_c$, the

ground state is ferromagnetically ordered and doubly degenerate, while for $g > g_c$ the system enters a disordered phase with a unique, even-parity ground state.

Applying the Jordan-Wigner transformation,

$$\sigma_\ell^x = 1 - 2c_\ell^\dagger c_\ell,$$
$$\sigma_\ell^z = -(c_\ell + c_\ell^\dagger) \prod_{p < \ell}(1 - 2c_p^\dagger c_p),$$

the even sector of the Hamiltonian can be rewritten in terms of the fermions as a quadratic form

$$H^+ = \sum_{\ell=1}^N \left[ g\left(c_\ell^\dagger c_\ell - \frac{1}{2}\right) - c_\ell^\dagger c_{\ell+1} + c_\ell c_{\ell+1} \right] + \text{h.c.}, \tag{S26}$$

with antiperiodic boundary conditions $c_{N+1} = -c_1$. The allowed pseudomomenta are

$$k = \pm \frac{1}{2}\frac{2\pi}{N}, \pm\frac{3}{2}\frac{2\pi}{N}, \cdots, \pm\frac{N-1}{2}\frac{2\pi}{N}. \tag{S27}$$

Introducing the Fourier transform,

$$c_\ell = \frac{e^{-i\pi/4}}{\sqrt{N}} \sum_k c_k e^{ik\ell}, \tag{S28}$$

and discarding constant terms, the Hamiltonian becomes

$$H^+ = \sum_k \left[ (g - \cos k)\left(c_k^\dagger c_k - c_{-k}c_{-k}^\dagger\right) + \sin k \left(c_k^\dagger c_{-k}^\dagger + c_{-k}c_k\right)\right]. \tag{S29}$$

Diagonalizing the above via a Bogoliubov transformation yields the energy spectrum

$$\epsilon_k = \pm 2\sqrt{(g - \cos k)^2 + \sin^2 k}, \tag{S30}$$

with the ground state obtained by filling all negative-energy modes. The excitations are described by

$$\eta_k = \cos(\varphi_k/2)c_k - \sin(\varphi_k/2)c_{-k}^\dagger,$$
$$\eta_k^\dagger = \cos(\varphi_k/2)c_k^\dagger - \sin(\varphi_k/2)c_{-k}, \tag{S31}$$

where $(\cos\varphi_k, \sin\varphi_k) = (2/\epsilon_k)(g - \cos k, \sin k)$. The Hamiltonian in the even sector then reads

$$H^+ = \sum_k \epsilon_k \left(\eta_k^\dagger \eta_k - \frac{1}{2}\right), \tag{S32}$$

with the Hilbert space restricted to states containing an even number of excitations. At the critical point $g = g_c = 1$, the dispersion reduces to

$$\epsilon_k = 4\left|\sin\frac{k}{2}\right|. \tag{S33}$$

## B. Dynamical structure factor

Given the complete spectrum $\{|n\rangle, E_n\}$, the dynamical structure factor of an operator $\phi$ is defined as [3]

$$S(k, \omega) = \frac{2\pi}{\mathcal{Z}V} \sum_{n,m} e^{-E_n/T} |\langle n|\phi(k)|m\rangle|^2 \, \delta(\omega - E_m + E_n). \tag{S34}$$

Physically, this quantity may be interpreted as a Gibbs-weighted average over initial states $|n\rangle$, multiplied by the Fermi's-golden-rule transition probability into final states $|m\rangle$ whose energy difference matches the frequency $\omega$.

*1. Evaluation of matrix elements*

We focus on the dynamical structure factor of the operator

$$X_{\text{tot}} \equiv \sum_\ell \sigma_\ell^x = \sum_\ell \left(1 - 2c_\ell^\dagger c_\ell\right) = \sum_k \left(1 - 2c_k^\dagger c_k\right), \tag{S35}$$

which is invariant under the $\mathbb{Z}_2$ symmetry. Expressing the fermion number operator in the Bogoliubov basis,

$$c_k^\dagger c_k = \left(\cos(\tfrac{\varphi_k}{2})\eta_k^\dagger - \sin(\tfrac{\varphi_k}{2})\eta_{-k}\right)\left(\cos(\tfrac{\varphi_k}{2})\eta_k - \sin(\tfrac{\varphi_k}{2})\eta_{-k}^\dagger\right), \tag{S36}$$

we see that $X_{\text{tot}}$ can only flip a pair of momentum modes $\{k, -k\}$. Energy conservation then requires $\omega = \pm 2\epsilon_k$.

The Fermi's-golden-rule matrix element between the ground state and a state with excitations $\{\eta_k^\dagger, \eta_{-k}^\dagger\}$ is

$$\left|\langle gs|X_{\text{tot}}\eta_k^\dagger \eta_{-k}^\dagger|gs\rangle\right|^2 = \left|\langle gs|(1 - 2c_k^\dagger c_k)\eta_k^\dagger \eta_{-k}^\dagger|gs\rangle\right|^2 \tag{S37}$$

$$= \left|\langle gs| - 2\left(-\sin(\tfrac{\varphi_k}{2})\eta_{-k}\right)\left(\cos(\tfrac{\varphi_k}{2})\eta_k\right)\eta_k^\dagger \eta_{-k}^\dagger|gs\rangle\right|^2$$

$$= \left|\sin\varphi_k \langle gs|\eta_{-k}\eta_k \eta_k^\dagger \eta_{-k}^\dagger|gs\rangle\right|^2$$

$$= \sin^2\varphi_k. \tag{S38}$$

Enforcing $\omega = \pm 2\epsilon_k$, we identify

$$\epsilon_k = \frac{|\omega|}{2} = 4|\sin(k/2)|. \tag{S39}$$

The momentum $q$ satisfying energy conservation is thus

$$q = q(\omega) \equiv 2\sin^{-1}\left(\sqrt{\frac{|\omega|}{8}}\right). \tag{S40}$$

Consequently,

$$\sin^2\varphi_k = \left(\frac{2}{\epsilon_k}\sin k\right)^2 = \left(\frac{2}{4\sin(k/2)}\sin k\right)^2 = \cos^2\left(\tfrac{k}{2}\right) = 1 - \sin^2\left(\tfrac{k}{2}\right) = 1 - \frac{|\omega|}{8}. \tag{S41}$$

We can now evaluate the dynamical structure factor step-by-step. For positive frequency $\omega > 0$, a nontrivial matrix element arises only when $X_{\text{tot}}$ acts on eigenstates without $\{q, -q\}$ excitations, denoted $n/\{q, -q\}$. These satisfy $\eta_q|n\rangle = \eta_{-q}|n\rangle = 0$. Hence,

$$S(\omega > 0) = \frac{2\pi}{\mathcal{Z}V}\sum_n\sum_k e^{-E_n/T}|\langle n|X_{\text{tot}}\eta_k^\dagger \eta_{-k}^\dagger|n\rangle|^2 \delta(\omega - 2\epsilon_k) \tag{S42}$$

$$= \frac{2\pi}{\mathcal{Z}V}\sum_{n/\{q,-q\}} e^{-E_n/T}\left(1 - \frac{\omega}{8}\right). \tag{S43}$$

For negative frequency $\omega < 0$, the relevant matrix element comes from eigenstates where the $\{q, -q\}$ modes are already occupied, denoted $n \supset \{q, -q\}$, which satisfy $\eta_q^\dagger|n\rangle = \eta_{-q}^\dagger|n\rangle = 0$. The result is

$$S(\omega < 0) = \frac{2\pi}{\mathcal{Z}V}\sum_{n \supset \{q,-q\}} e^{-E_n/T}\left(1 - \frac{|\omega|}{8}\right) \tag{S44}$$

$$= \frac{2\pi}{\mathcal{Z}V}\sum_{n/\{q,-q\}} e^{-(E_n+|\omega|)/T}\left(1 - \frac{|\omega|}{8}\right)$$

$$= e^{-|\omega|/T}S(-\omega), \tag{S45}$$

where we used $\epsilon_{-q} + \epsilon_q = 2\epsilon_q = \omega$.

### 2. Partition function

As noted in deriving Eq. (S43), it is necessary to account for eigenstates that do not contain excitations at momenta $q$ and $-q$. While $X_{\text{tot}}$ acting on the ground state only generates $\{k, -k\}$ pairs of excitations, one must carefully recognize that these do not exhaust all low-energy excitations of the Hamiltonian. For example, the state

$$\eta^\dagger_{-\frac{\pi}{101}} \eta^\dagger_{-\frac{2\pi}{101}} \eta^\dagger_{-\frac{3\pi}{101}} \eta^\dagger_{\frac{6\pi}{101}} |gs\rangle$$

is a valid eigenstate with zero net momentum and energy below the single-particle bandwidth, yet it cannot be generated from the ground state by $X_{\text{tot}}$.

If we were to restrict our attention solely to $\{k, -k\}$ excitations, neglecting all other possibilities, the result would be $S(\omega) \sim 1/[1 + e^{-\omega/T}]$. However, once the full set of allowed low-energy excitations is included, a different expression emerges.

For a given $\omega$ and the corresponding $q = q(\omega)$, the partition function can be decomposed as

$$\mathcal{Z} = \sum_{\substack{n \\ \text{even}}} e^{-E_n/T} \tag{S46}$$

$$= \sum_{\{q,-q\} \notin n} e^{-E_n/T} + \sum_{\{q,-q\} \in n} e^{-E_n/T} + \sum_{q \in n, -q \notin n} e^{-E_n/T} + \sum_{-q \in n, q \notin n} e^{-E_n/T}$$

$$= \sum_{\substack{n/\{q,-q\} \\ \text{even}}} e^{-E_n/T} + \sum_{\substack{n/\{q,-q\} \\ \text{even}}} e^{-(E_n+\omega)/T} + \sum_{\substack{n/\{q,-q\} \\ \text{odd}}} e^{-(E_n+\omega/2)/T} + \sum_{\substack{n/\{q,-q\} \\ \text{odd}}} e^{-(E_n+\omega/2)/T}$$

$$= (1 + e^{-\omega/T}) \left( \sum_{\substack{n/\{q,-q\} \\ \text{even}}} e^{-E_n/T} \right) + 2e^{-\omega/2T} \left( \sum_{\substack{n/\{q,-q\} \\ \text{odd}}} e^{-E_n/T} \right)$$

$$\equiv (1 + e^{-\omega/T}) \, \mathcal{Z}_{\text{even}}(q) + 2e^{-\omega/2T} \, \mathcal{Z}_{\text{odd}}(q), \tag{S47}$$

where the "even" and "odd" subscripts denote the parity of the number of excitations. Here, $\mathcal{Z}_{\text{even}}(q)$ and $\mathcal{Z}_{\text{odd}}(q)$ are partition functions restricted to states with an even or odd number of excitations, respectively, excluding the $\{q, -q\}$ modes.

At the critical point, where the spectrum is gapless, one expects $\mathcal{Z}_{\text{even}}(q) = \mathcal{Z}_{\text{odd}}(q)$ in the thermodynamic limit.

### 3. Absence of a resonance

Substituting Eq. (S47) into Eq. (S43) results in

$$S(\omega > 0) = \frac{2\pi}{L} \frac{\left(1 - \frac{\omega}{8}\right) \mathcal{Z}_{\text{even}}(q)}{(1 + e^{-\omega/T})\mathcal{Z}_{\text{even}}(q) + 2e^{-\omega/2T}\mathcal{Z}_{\text{odd}}(q)} \tag{S48}$$

$$= \frac{2\pi}{L} \frac{1 - \frac{\omega}{8}}{(1 + e^{-\omega/T}) + 2e^{-\omega/2T} \, \mathcal{Z}_{\text{odd}}(q)/\mathcal{Z}_{\text{even}}(q)}. \tag{S49}$$

At criticality, where $\mathcal{Z}_{\text{even}}(q) = \mathcal{Z}_{\text{odd}}(q)$, this simplifies to

$$S(\omega > 0) = \frac{2\pi}{L} \left(1 - \frac{\omega}{8}\right) \frac{1}{(1 + e^{-\omega/2T})^2}. \tag{S50}$$

For negative frequencies, one finds

$$S(\omega < 0) = e^{-|\omega|/T} S(-\omega) = \frac{2\pi}{L} \left(1 + \frac{\omega}{8}\right) \frac{e^{\omega/T}}{(1 + e^{\omega/2T})^2} \tag{S51}$$

$$= \frac{2\pi}{L} \left(1 + \frac{\omega}{8}\right) \frac{1}{(1 + e^{-\omega/2T})^2}. \tag{S52}$$

Combining Eqs. (S50) and S52, the dynamical structure factor at the critical point reduces to

$$S(\omega) = \frac{2\pi}{L} \frac{1}{(1 + e^{-\omega/2T})^2},$$

(S53)

which is valid for frequencies $\omega$ much smaller than the single-particle bandwidth.

Thus, the dynamical structure factor $S(\omega)$ exhibits no resonance peak. Instead, it increases smoothly from $S(\omega \to -\infty) = 0$ to $S(\omega \to \infty) = 1$. This behavior confirms the absence of an amplitude mode at the $(1+1)$D Ising critical point *at the level of linear response*, in contrast to higher-dimensional $\mathcal{O}(\mathcal{N} \geq 2)$ critical theories.

### SIII. Nonequilibrium spectroscopy of the transverse-field Ising model

Going beyond linear-response theory, we now investigate the possible existence of an amplitude mode at the $(1+1)$D Ising QCP using quench spectroscopy, in the same spirit as the the analysis of the XXZ model presented in the main text.

#### A. Quench dynamics

Motivated by the fact that the oscillations in the XXZ model were revealed by a four-fermion probe, we consider an analogous four-fermion setup in the transverse-field Ising model. Here, the natural fermions are Majoranas, and the term associated with $\lambda$ in Eq. (S54) maps to a product of four consecutive Majorana operators, as we explain in the next section.

Our protocol is defined as follows. Starting with the ferromagnetic TFIM—with open boundary conditions—in a longitudinal field and additional Ising interactions,

$$H = -\sum_\ell \left(\sigma_\ell^z \sigma_{\ell+1}^z + g\sigma_\ell^x\right) + h\sum_\ell \sigma_\ell^z - \lambda \sum_\ell \left(\sigma_\ell^x \sigma_{\ell+1}^x + \sigma_\ell^z \sigma_{\ell+2}^z\right),$$

(S54)

we consider quenches to the gapless QCP located at $g = 1$, $h = \lambda = 0$.

First, we study an *interaction* quench wherein an initial state is prepared in the ground state at $\lambda = \lambda_i$ (with $g = 1$, $h = 0$ fixed) and the interaction term is quenched thereafter to $\lambda = 0$. We track the post-quench dynamics numerically using TEBD for a spin chain with $N = 100$ sites, and probe the composite operator $\sum_\ell \langle S_\ell^x S_{\ell+1}^x + S_\ell^z S_{\ell+2}^z \rangle / N$, which corresponds to a sum over terms with four consecutive Majorana fermions. This choice is designed to parallel the four-fermion probes used for the XXZ model. The resulting dynamics are shown in Fig. S3(a) and exhibit underdamped early-time oscillations.

We also explore a *field* quench that uses a different control knob in Eq. (S54): the system is initialized in the ground state at $h = h_i$ (with $g = 1$, $\lambda = 0$ fixed) and the longitudinal field is quenched to zero. Note that since $g = 1$, the model is always gapless at the end of the quench. In this case, we monitor two $\mathbb{Z}_2$-symmetric observables, namely, the transverse magnetization $\sum_\ell \langle S_\ell^x \rangle / N$, as well as the nearest-neighbor correlator $\sum_\ell \langle S_\ell^x S_{\ell+1}^x \rangle / N$. As shown in Figs. S3(b,c), we again observe out-of-equilibrium oscillations akin to the XXZ model.

#### B. Majorana representation

Starting from Jordan–Wigner fermions $c_\ell$, we define two Majoranas per site

$$\gamma_{2\ell-1} \equiv c_\ell + c_\ell^\dagger, \qquad \gamma_{2\ell} \equiv i\left(c_\ell^\dagger - c_\ell\right), \qquad \{\gamma_m, \gamma_n\} = 2\delta_{mn}.$$

(S55)

With the standard Jordan-Wigner mapping for the TFIM,

$$\sigma_\ell^x = 1 - 2c_\ell^\dagger c_\ell = -i\,\gamma_{2\ell-1}\gamma_{2\ell}, \qquad \sigma_\ell^z \sigma_{\ell+1}^z = i\,\gamma_{2\ell}\gamma_{2\ell+1}.$$

(S56)

Therefore,

$$\sum_\ell \sigma_\ell^x \sigma_{\ell+1}^x = -\sum_\ell \gamma_{2\ell-1}\gamma_{2\ell}\gamma_{2\ell+1}\gamma_{2\ell+2},$$

(S57)

$$\sum_\ell \sigma_\ell^z \sigma_{\ell+2}^z = -\sum_\ell \gamma_{2\ell}\gamma_{2\ell+1}\gamma_{2\ell+2}\gamma_{2\ell+3}.$$

(S58)

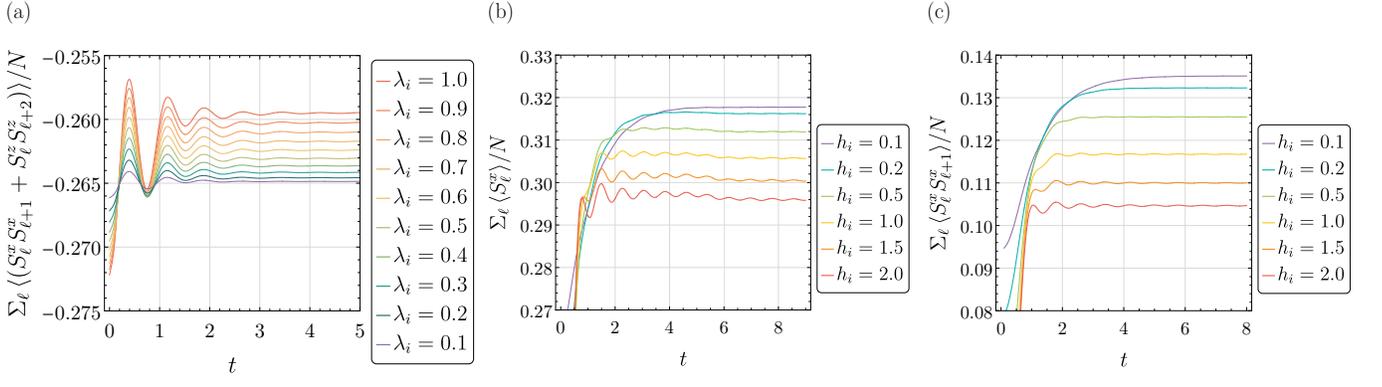

FIG. S3. (a) Time evolution of $\sum_{\ell}\langle S_{\ell}^{x}S_{\ell+1}^{x}+S_{\ell}^{z}S_{\ell+2}^{z}\rangle/N$ in the $(1+1)$D TFIM (as computed with TEBD for a 100-site system), after a quench of the interaction from $\lambda=\lambda_i$ to $\lambda=0$ with $g=1$, $h=0$ throughout. (b) Dynamics of the $\mathbb{Z}_2$-symmetric probes $\sum_{\ell}\langle S_{\ell}^{z}\rangle/N$ and (c) $\sum_{\ell}\langle S_{\ell}^{x}S_{\ell+1}^{x}\rangle/N$ following a quench of the longitudinal field from $h=h_i$ to $h=0$ while setting $g=1$, $\lambda=0$.

These are terms with four consecutive Majorana operators, starting on odd and even Majorana sites, respectively. Combining them, we define

$$\mathcal{O}_4 \equiv -\sum_{\ell}\Big(\sigma_{\ell}^{x}\sigma_{\ell+1}^{x}+\sigma_{\ell}^{z}\sigma_{\ell+2}^{z}\Big)=\sum_{m}\gamma_m\gamma_{m+1}\gamma_{m+2}\gamma_{m+3}, \tag{S59}$$

which is a translationally invariant sum over sets of four consecutive Majorana operators with unit lattice spacing in the Majorana chain. This is directly analogous to the four-fermion term relevant in the XXZ case, and motivates our inclusion of these interactions as the operator being quenched in the Hamiltonian (S54).

In the even-parity sector, the TFIM Hamiltonian becomes

$$H^{+} = \frac{i}{2}\sum_{\ell=1}^{N}\Big(2g\,\gamma_{2\ell-1}\gamma_{2\ell}+2\,\gamma_{2\ell}\gamma_{2\ell+1}\Big), \tag{S60}$$

with antiperiodic boundary conditions $\gamma_{m+2N}=-\gamma_m$ after relabeling Majorana sites $m=1,\ldots,2N$. At criticality $g=1$, this reduces to a uniform nearest-neighbor chain with one-Majorana-site translation symmetry,

$$H^{+} = i\sum_{m=1}^{2N}\gamma_m\gamma_{m+1}. \tag{S61}$$

Antiperiodic boundary conditions set the allowed momenta as

$$q=\pm\frac{\pi}{2N},\ \pm3\frac{\pi}{2N},\ \ldots,\ \pm(2N-1)\frac{\pi}{2N}, \tag{S62}$$

over the range $q\in(-\pi,\pi)$. We use the Fourier transform

$$\gamma_m=\frac{1}{\sqrt{2N}}\sum_{q}e^{iqm}\,\gamma_q,\qquad \gamma_{-q}=\gamma_q^{\dagger},\qquad \{\gamma_q,\gamma_{q'}\}=2\,\delta_{q,-q'} \tag{S63}$$

where the momenta $q$ run over the range given in Eq. (S62). Substituting Eq. (S63) into Eq. (S61) gives

$$\begin{aligned}
H^{+}&=i\sum_{m=1}^{2N}\gamma_m\gamma_{m+1}=i\sum_{q,q'}\delta_{q,-q'}\,e^{iq'}\,\gamma_q\gamma_{q'}\\
&=i\sum_{0<q<\pi}\Big(e^{-iq}\gamma_q\gamma_{-q}+e^{iq}\gamma_{-q}\gamma_q\Big)=\sum_{0<q<\pi}\Big(4\cos q+\sin q\,(\gamma_q\gamma_{-q}-\gamma_{-q}\gamma_q)\Big)\\
&=\sum_{0<q<\pi}\sin q\,(\gamma_q\gamma_{-q}-\gamma_{-q}\gamma_q). \tag{S64}
\end{aligned}$$

It is convenient to define, for $q \in (0, \pi)$,

$$f_q := \frac{\gamma_{-q}}{\sqrt{2}}, \qquad f_q^\dagger = \frac{\gamma_q}{\sqrt{2}}, \qquad \{f_q, f_{q'}\} = \{f_q^\dagger, f_{q'}^\dagger\} = 0, \quad \{f_q, f_{q'}^\dagger\} = \delta_{q,q'}. \tag{S65}$$

In terms of these complex fermions,

$$H^+ = \sum_{0 < q < \pi} 2 \sin q \left(2 f_q^\dagger f_q - 1\right), \tag{S66}$$

so each pair $(\gamma_q, \gamma_{-q})$ forms an independent two-level system with energies $\pm 2|\sin q|$. There are $N$ independent $q$-pairs, resulting in a total Hilbert-space dimension of $2^N$.

The ground state $|\text{GS}\rangle$ is the vacuum of the $f_q$ modes,

$$f_q |\text{GS}\rangle = 0 \quad \text{for all } q \in (0, \pi), \tag{S67}$$

and, equivalently, $\gamma_{-q} |\text{GS}\rangle = 0$ for all $q \in (0, \pi)$ since $f_q = \gamma_{-q}/\sqrt{2}$. The ground-state energy is

$$E_{\text{GS}} = -2 \sum_{0 < q < \pi} \sin q = -2 \sum_{n=1}^{N} \sin\left(\frac{(2n-1)\pi}{2N}\right). \tag{S68}$$

### C. Four-Majorana operator

We consider the operator

$$\mathcal{O}_4 \equiv \sum_m \gamma_m \, \gamma_{m+1} \, \gamma_{m+2} \, \gamma_{m+3}. \tag{S69}$$

Using the Fourier transform in Eq. (S63), we obtain

$$\begin{aligned}
\mathcal{O}_4 &= \frac{1}{(2N)^2} \sum_m \sum_{k_1, k_2, k_3, k_4} e^{i\,[k_1 m + k_2(m+1) + k_3(m+2) + k_4(m+3)]} \gamma_{k_1} \gamma_{k_2} \gamma_{k_3} \gamma_{k_4} \\
&= \frac{1}{2N} \sum_{k_1, k_2, k_3, k_4} \delta_{2\pi}(k_1 + k_2 + k_3 + k_4) \, e^{i\,(k_2 + 2k_3 + 3k_4)} \gamma_{k_1} \gamma_{k_2} \gamma_{k_3} \gamma_{k_4}.
\end{aligned} \tag{S70}$$

Only the totally antisymmetric part in $(k_1, k_2, k_3, k_4)$ survives because the $\gamma_k$ anticommute. Let us define

$$\begin{aligned}
F^{\text{A}}(k_1, k_2, k_3, k_4) &\equiv \frac{1}{4!} \sum_{\pi \in S_4} \text{sgn}(\pi) \, e^{i\,(k_{\pi(2)} + 2k_{\pi(3)} + 3k_{\pi(4)})} = \frac{1}{4!} \prod_{1 \le a < b \le 4} \left(e^{ik_b} - e^{ik_a}\right) \\
&= \frac{(2i)^6}{4!} \, e^{\frac{3i}{2} \sum_{a=1}^{4} k_a} \prod_{1 \le a < b \le 4} \sin \frac{k_b - k_a}{2}.
\end{aligned} \tag{S71}$$

Thus,

$$\mathcal{O}_4 = \frac{1}{2N} \sum_{k_1, k_2, k_3, k_4} \delta_{2\pi}(\kappa) \, F^{\text{A}}(k_1, k_2, k_3, k_4) \, \gamma_{k_1} \gamma_{k_2} \gamma_{k_3} \gamma_{k_4}, \qquad \kappa \equiv k_1 + k_2 + k_3 + k_4. \tag{S72}$$

As $\gamma_k$ with negative momentum annihilates the ground state ($\gamma_{k<0} |\text{GS}\rangle = 0$; equivalently, $\gamma_{-q} |\text{GS}\rangle = 0$ for $q \in (0, \pi)$), the only nonzero matrix elements from $|\text{GS}\rangle$ arise when all $k_1, k_2, k_3, k_4 \in (0, \pi)$. In that case, $\delta_{2\pi}(\kappa)$ enforces $\kappa = 2\pi$.

We define the four-Majorana excited state using Eq. (S65) (with $k_j \in (0, \pi)$),

$$|k_1, k_2, k_3, k_4\rangle \; = \; f_{k_1}^\dagger f_{k_2}^\dagger f_{k_3}^\dagger f_{k_4}^\dagger |\text{GS}\rangle, \tag{S73}$$

Then, the corresponding matrix element is

$$\langle k_1, k_2, k_3, k_4 | \, \mathcal{O}_4 \, |\text{GS}\rangle = \frac{4!}{2N} \, (\sqrt{2})^4 \, F^{\text{A}}(k_1, k_2, k_3, k_4) \, \delta_{2\pi}(\kappa). \tag{S74}$$

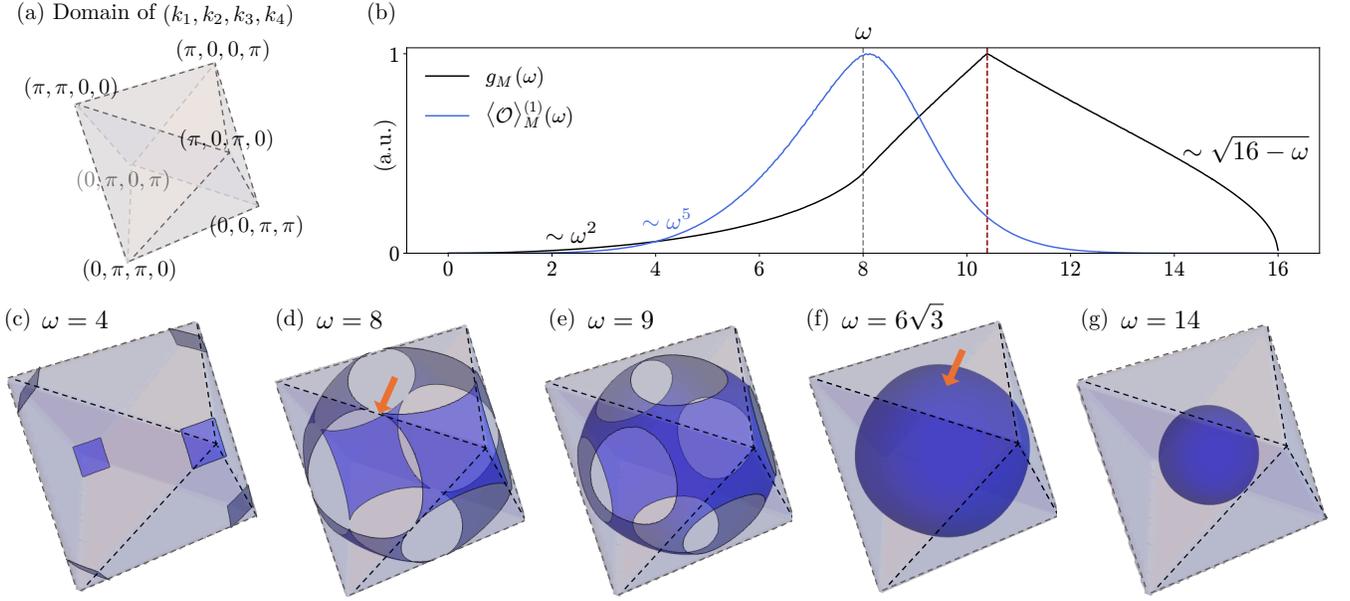

FIG. S4. (a) Domain of $(k_1, k_2, k_3, k_4)$ with each $k_a \in (0, \pi)$ and total momentum $\sum_a k_a = 2\pi$. The allowed set is a regular octahedron; each vertex corresponds to two momenta at $\pi$ and two at 0. (b) Density of states $g_M(\omega)$ and the perturbative Fourier component $\langle \mathcal{O} \rangle_M^{(1)}(\omega)$ versus $\omega/J$. Gray and red dotted lines mark the singular frequencies. (c–g) Constant-energy surfaces (blue) inside the octahedral domain for selected $\omega/J$: (c) 4, (d) 8, (e) 9, (f) $6\sqrt{3}$, and (g) 14. The surface touches an edge at $\omega/J = 8$ (d) and a face at $\omega/J = 6\sqrt{3}$ (f), indicated by orange arrows; these contacts produce Van Hove singularities shown in panel (b).

## D. Density of states and singularities

As derived in Eq. (S74), the selection rule for the four-Majorana excitation created by the operator $\mathcal{O}_4$ enforces momentum conservation, so $\kappa = \sum_a k_a = 2\pi$. Unlike the complex-fermion case in the XX model with two holes and two particles (that are only pairwise symmetric), the four Majoranas are indistinguishable and fully symmetric under all permutations. With one linear constraint, the four momenta can be represented in a three-dimensional space, with each $k_a \in (0, \pi)$. By symmetry, the allowed region is a regular octahedron, as shown in Fig. S4(a). Each vertex corresponds to two momenta at 0 and two at $\pi$, and the center is at $(k_1, k_2, k_3, k_4) = (\frac{\pi}{2}, \frac{\pi}{2}, \frac{\pi}{2}, \frac{\pi}{2})$.

As in the XX model, Fig. S4(b) plots the normalized DOS $g_M(\omega)$ (black) and $\langle \mathcal{O} \rangle_M^{(1)}(\omega)$ (blue) from Eq. (S7), while Figs. S4(c-g) show the corresponding constant-energy surfaces. As $\omega$ decreases from the top of the band $\omega_{T,M} = 16J$, the surface grows from the central point into an ellipsoid-like shape [Fig. S4(g)] until it first touches a face of the octahedron [Fig. S4(f)]. This contact produces a Van Hove singularity and a cusp in $g_M(\omega)$ at $\omega_{F,M}/J = 6\sqrt{3}$. A second morphological change occurs at $\omega_{E,M}/J = 8$ [Figs. S4(e) to (c)] when the surface hits an edge, which produces another Van Hove singularity. There are also trivial singularities at the bottom $\omega_{B,M} = 0$ and top $\omega_{T,M} = 16J$ of the band.

While the DOS structure resembles that of free complex fermions in the XX model, the matrix elements differ and strongly affect the Fourier component. Since the four Majoranas are indistinguishable, the matrix element is fully symmetric and vanishes when the constant-energy surface touches faces, edges, or vertices. Let $\delta\omega$ denote the detuning from a singular frequency; the matrix elements exhibit a different scaling with $\delta\omega$ near the singularity as compared to the XX model. Recall that the dispersion is $E(\{k\}) = 4\sum_a \sin k_a$. The leading scalings of the matrix element near singular points inside the octahedron are:

- Near $\omega_{B,M}$ (bottom of the band): Two momenta approach $\pi$ (say $k_1, k_2$) and two approach 0 (say $k_3, k_4$). Then, $|\delta\omega| \sim |\delta k|$, and the matrix element goes as

$$\sin\left(\frac{k_1 - k_2}{2}\right) \sin\left(\frac{k_3 - k_4}{2}\right) \sim (\delta k)^2 \sim (\delta\omega)^2. \tag{S75}$$

- Near $\omega_{E,M}$ (edge touch): Two momenta are close to $\pi/2$ (say $k_1, k_2$), with the other two near $\pi$ and 0 (say $k_3$

and $k_4$). Here, $\delta\omega = \mathcal{O}(\delta k_1^2) + \mathcal{O}(\delta k_2^2)$, so $\delta k_{1,2} \sim |\delta\omega|^{1/2}$, and the matrix element scales as

$$\sin\left(\frac{k_1 - k_2}{2}\right) \sim |\delta\omega|^{1/2}. \tag{S76}$$

- Near $\omega_{F,M}$ (face touch): Three momenta are close to $2\pi/3$ (say $k_1, k_2, k_3$) and the remaining one is near $0$ or $\pi$ (say $k_4$). Then, $|\delta\omega| \sim |\delta k|$, and the matrix element involves

$$\sin\left(\frac{k_1 - k_2}{2}\right) \sin\left(\frac{k_2 - k_3}{2}\right) \sin\left(\frac{k_3 - k_1}{2}\right) \sim |\delta k|^3 \sim |\delta\omega|^3. \tag{S77}$$

- Near $\omega_{T,M}$ (top of the band): All momenta satisfy $k_a = \pi/2 + \delta k_a$ with $|\delta k_a| \ll 1$. Since

$$\delta\omega = O(\delta k_1^2 + \delta k_2^2 + \delta k_3^2 + \delta k_4^2), \tag{S78}$$

typically, $|\delta k| \sim |\delta\omega|^{1/2}$. The matrix element includes products of pairwise differences, so

$$\prod_{i<j} \sin\left(\frac{k_i - k_j}{2}\right) \sim (\delta k)^6 \sim |\delta\omega|^3. \tag{S79}$$

### SIV.  Bethe ansatz equations and solutions

In this section, we review the Bethe ansatz equations for the XXZ spin chain in the easy-axis regime and summarize the associated excitation families, following Chapters 4 and 9 of Ref. 4.

#### A.  Easy-axis XXZ spin chain

As discussed in the main text, we consider the ferromagnetic XXZ spin chain, which is unitarily related to the antiferromagnetic XXZ chain with anisotropy inverted, i.e., $\Delta_{\mathrm{FM}} = -\Delta$. The Hamiltonian is

$$H = -J \sum_{\ell=1}^{N} \left( S_\ell^x S_{\ell+1}^x + S_\ell^y S_{\ell+1}^y + \Delta_{\mathrm{FM}} S_\ell^z S_{\ell+1}^z \right); \quad J > 0, \tag{S80}$$

with periodic boundary conditions imposed, $S_{N+1} = S_1$. Taking the fully polarized state as the vacuum, we work in the sector with $M$ spin flips, characterized by $M$ quasimomenta $\{k_j\}$. Once these quasimomenta are determined, the corresponding energy is

$$E = E_0 + \sum_{j=1}^{M} J\left(\Delta_{\mathrm{FM}} - \cos k_j\right), \tag{S81}$$

where $E_0 = -J\,\Delta_{\mathrm{FM}}\,N/4$ is the vacuum energy.

For $-1 < \Delta_{\mathrm{FM}} < 1$, it is convenient to introduce $\Gamma \equiv \arccos(-\Delta_{\mathrm{FM}})$, which lies in the interval $0 < \Gamma < \pi$. Defining rapidities $x_j$ by

$$\exp(ik_j) \equiv -\frac{\sinh\frac{\Gamma}{2}(x_j + i)}{\sinh\frac{\Gamma}{2}(x_j - i)}, \tag{S82}$$

the periodic boundary conditions lead to the Bethe equations

$$\left[\frac{\sinh\frac{\Gamma}{2}(x_j + i)}{\sinh\frac{\Gamma}{2}(x_j - i)}\right]^N = \prod_{\ell\neq j} \frac{\sinh\frac{\Gamma}{2}(x_j - x_\ell + 2i)}{\sinh\frac{\Gamma}{2}(x_j - x_\ell - 2i)}. \tag{S83}$$

Taking the logarithm then gives

$$N\,\theta(x_j, 1) = 2\pi I_j + \sum_{\ell=1}^{M} \theta(x_j - x_\ell, 2), \tag{S84}$$

where the Bethe quantum numbers (BQNs) $I_j$ are distinct integers when $M$ is odd and distinct half-odd integers when $M$ is even. Here, we have defined the function

$$\theta(x, n) \equiv 2\tan^{-1}\left(\frac{\tanh\frac{\Gamma x}{2}}{\tan\frac{n\Gamma}{2}}\right). \tag{S85}$$

Eq. (S82) is equivalent to the statement that $k_j \equiv \theta(x_j, 1)$. Since $\theta(\cdot, n)$ is an odd function, summing Eq. (S84) over all $j$ yields the total momentum

$$K_{\text{tot}} = \sum_j k_j = \frac{2\pi}{N}\sum_j I_j. \tag{S86}$$

The energy can also be expressed in terms of the rapidities as

$$E(x_j) - E_0 = -\sum_j \frac{J\sin^2\Gamma}{\cosh\Gamma x_j - \cos\Gamma} = -\frac{2\pi J\sin\Gamma}{\Gamma}\sum_j a(x_j, 1), \tag{S87}$$

with kernel

$$a(x, n) \equiv \frac{1}{2\pi}\frac{d}{dx}\theta(x, n) = \frac{1}{2\pi}\frac{\Gamma\sin(n\Gamma)}{\cosh\Gamma x - \cos(n\Gamma)}. \tag{S88}$$

For a fixed magnetization sector with $M$ spin flips, the ground-state BQNs form a contiguous set centered at zero:

$$\{I_j\}_{\text{GS}} = \left\{-\frac{M-1}{2}, -\frac{M-3}{2}, \ldots, \frac{M-1}{2}\right\}. \tag{S89}$$

These are, of course, integers when $M$ is odd and half-odd integers when $M$ is even.

For later use, it is convenient to also consider rapidities shifted by a purely imaginary amount $ip_0$, with $p_0 \equiv \pi/\Gamma$, i.e., $x_j = y_j + ip_0$ with $y_j \in \mathbb{R}$. In this case,

$$k_j = \pi - \theta(y_j, p_0 - 1), \qquad E(y_j) - E_0 = \frac{2\pi J\sin\Gamma}{\Gamma}a(y_j, p_0 - 1). \tag{S90}$$

Similarly, a two-string solution sharing a real part $z_j$ has energy

$$E(z_j) - E_0 = -J\frac{\sin\Gamma\sin 2\Gamma}{\cosh\Gamma z_j - \cos\Gamma} = -\frac{2\pi J\sin\Gamma}{\Gamma}a(z_j, 2). \tag{S91}$$

### B. Valid excitations

In the previous section, we described the ground state of the easy-axis XXZ spin chain, characterized by real rapidities $x \in \mathbb{R}$. We now discuss the admissible excitations and their representation.

As in the ground state, excitations can involve real rapidities. In addition, complex rapidities grouped into strings are allowed. An $n$-string consists of $n$ rapidities sharing a common real part and representing a bound state of $n$ spins. In the gapless XY-like regime with $|\Delta_{\text{FM}}| < 1$, two distinct types of strings arise, distinguished by the position of their centers: one type lies on the real axis, while the other is shifted by a purely imaginary offset $ip_0$, with $p_0 \equiv \pi/\Gamma$. The rapidities in an $n$-string take the form

$$(\nu = +1) \quad x_j = \alpha + (n + 1 - 2j)i, \tag{S92}$$

$$(\nu = -1) \quad x_j = \alpha + (n + 1 - 2j)i + p_0 i, \tag{S93}$$

where $j \in \{1, \ldots, n\}$, $\alpha \in \mathbb{R}$ is the real part of the string center, and $\nu = \pm 1$ labels the string type.

The admissible string lengths are restricted by the Takahashi–Suzuki conditions [5–7]:

$$2\sum_{j=1}^{n-1}\lfloor j\Gamma/\pi\rfloor = (n-1)\lfloor(n-1)\Gamma/\pi\rfloor, \tag{S94}$$

$$v\sin((n-1)\Gamma) \geq 0. \tag{S95}$$

where $\lfloor \cdot \rfloor$ denotes the floor function.

As explained in the previous section, the ground state consists of $M = N/2$ real rapidities ($\nu = +1$). Removing one real rapidity changes the number of rapidities to $M = N/2 \to N/2 - 1$, which flips the BQN grid between integer and half-odd-integer labels and reindexes the entire sea. Consequently, the "single-removal" configuration is naturally described as two holes—i.e., a two-*spinon* excitation. By contrast, a single $\nu = -1$ length-1 string is a one-particle mode with a single well-defined momentum; it does not fractionalize into two particles. We therefore refer to it as a *spin-wave* excitation.

In this work, we focus on excited states that remain charge neutral and involve at most four momentum degrees of freedom. This restriction is motivated by the free-fermion limit $\Delta_{\mathrm{FM}} = 0$, where the operator $\mathcal{O}^{zz} = \sum_i S_i^z S_{i+1}^z$ generates two-hole two-particle excitations subject to overall momentum conservation.

At $\Delta_{\mathrm{FM}} = 0$, only length-one rapidities are present, for both types $\nu = \pm 1$. The two-hole two-particle sector can then be constructed by removing two real rapidities of type $\nu = +1$ (spinons, or psinons when $S^z \neq 0$) and inserting two complex rapidities of type $\nu = -1$ (antispinons when $S^z \neq 0$). This naturally raises the question of how the two-spinon two-antispinon configuration generalizes to finite $\Delta_{\mathrm{FM}}$.

For $\Delta_{\mathrm{FM}} < 0$ (equivalently, $\Delta > 0$, the case of interest here), not all two-spinon two-antispinon states remain admissible. When complex rapidities of type $\nu = -1$ are present, we find numerically that the corresponding Bethe quantum numbers (BQNs) are confined to a narrow interval around zero. Specifically, for $m$ complex rapidities of type $\nu = -1$, the BQNs consistently take the form

$$\{I_c\} = \left\{ -\frac{m/2 - 1}{2}, -\frac{m/2 - 3}{2}, \cdots, \frac{m/2 - 1}{2} \right\}, \tag{S96}$$

while the BQNs of the remaining real rapidities remain unconstrained. We also numerically observe the same near-zero restriction for 2-string solutions, which we will consider later. Enforcing zero total momentum removes one degree of freedom, leaving only a single free parameter in the two-spinon two-antispinon family (two hole momenta minus one momentum-conservation constraint). This restricted phase space is insufficient to capture the collective oscillations observed at nonzero $\Delta_{\mathrm{FM}}$ (see Sec. IV D below).

We therefore turn to excitation families with four genuine momentum degrees of freedom, subject to a single momentum-conservation constraint, by considering more general string contents. Guided by TEBD simulations, which indicate that the oscillation persists smoothly as $\Delta_{\mathrm{FM}}$ is tuned away from zero, we focus on configurations that evolve continuously with the anisotropy, namely mixtures of one-strings and two-strings. From Eqs. (S94)–(S95), length-one rapidities are allowed for both string types $\nu = \pm 1$, while length-two strings are allowed only for type $\nu = +1$. In the next section, we present the Bethe ansatz equations for these mixed configurations, which involve three types of rapidities.

## C. Solving the Bethe ansatz equations

We now consider a generalized form of the Bethe equations that accommodates states with three distinct species of rapidities. For clarity, we denote them by the subscripts $r$, $c$, and $s$:

(i) Real length-one rapidities $x_j$ with Bethe quantum numbers (BQNs) $I_{r,j}$.

(ii) Complex length-one rapidities $x_{c,j} = y_j + ip_0$ with $y_j \in \mathbb{R}$ and BQNs $I_{c,j}$.

(iii) Length-two strings with real centers $z_j \in \mathbb{R}$ and BQNs $I_{s,j}$.

Here the subscript '$r$' corresponds to $\nu = +1$ length-one rapidities, '$c$' to $\nu = -1$ length-one rapidities, and '$s$' to $\nu = +1$ length-two strings. Let $M_r$, $M_c$, and $M_s$ denote the number of rapidities of each species. The total number of spin flips is then

$$M = M_r + M_c + 2M_s = N/2. \tag{S97}$$

Within each species ($r$, $c$, $s$), the BQNs $I_{\alpha,j}$ must be distinct. For each type $\alpha \in \{r, c, s\}$, the BQNs are integers when $M_\alpha$ is odd and half-odd integers when $M_\alpha$ is even.

Using $\theta(x, n)$ as defined in Eq. (S85), the coupled Bethe equations between the real variables $x_j$, $y_j$, and $z_j$ are

$$N\,\theta(x_j, 1) = 2\pi I_{r,j} + \sum_{\substack{\ell \in r \\ \ell \neq j}} \theta(x_j - x_\ell, 2) - \sum_{\ell \in c} \theta(x_j - y_\ell, p_0 - 2) + \sum_{\ell \in s} \big[\theta(x_j - z_\ell, 1) + \theta(x_j - z_\ell, 3)\big], \tag{S98}$$

$$-N\,\theta(y_j, p_0 - 1) = 2\pi I_{c,j} - \sum_{\ell \in r} \theta(y_j - x_\ell, p_0 - 2) + \sum_{\substack{\ell \in c \\ \ell \neq j}} \theta(y_j - y_\ell, 2) - \sum_{\ell \in s} \big[\theta(y_j - z_\ell, p_0 - 1) + \theta(y_j - z_\ell, p_0 - 3)\big], \tag{S99}$$

$$N\,\theta(z_j, 2) = 2\pi I_{s,j} + \sum_{\ell \in r} \big[\theta(z_j - x_\ell, 1) + \theta(z_j - x_\ell, 3)\big] - \sum_{\ell \in c} \big[\theta(z_j - y_\ell, p_0 - 1) + \theta(z_j - y_\ell, p_0 - 3)\big]$$
$$+ \sum_{\substack{\ell \in s \\ \ell \neq j}} \big[2\,\theta(z_j - z_\ell, 2) + \theta(z_j - z_\ell, 4)\big]. \tag{S100}$$

In each equation, the first term encodes the BQN assignment, while the remaining sums describe two-body scattering between rapidities of type ($r$), ($c$), and ($s$). The sums $\sum_{\ell \in r}$, $\sum_{\ell \in c}$, and $\sum_{\ell \in s}$ extend over all rapidities of the indicated species. These equations reduce to the standard one-species form [Eq. (S84)] when only type ($r$) rapidities are present.

Once the excitation configuration is specified and BQNs are assigned, the resulting $M$ coupled nonlinear Bethe equations are solved numerically using `scipy.optimize.root` (via the HYBR method) [8]. Whenever the solver converges, we record the corresponding energy. To match the TEBD simulations, we fix the system size to $N = 100$ sites, so that $M = 50$.

### D. Excitations with four momentum degrees of freedom

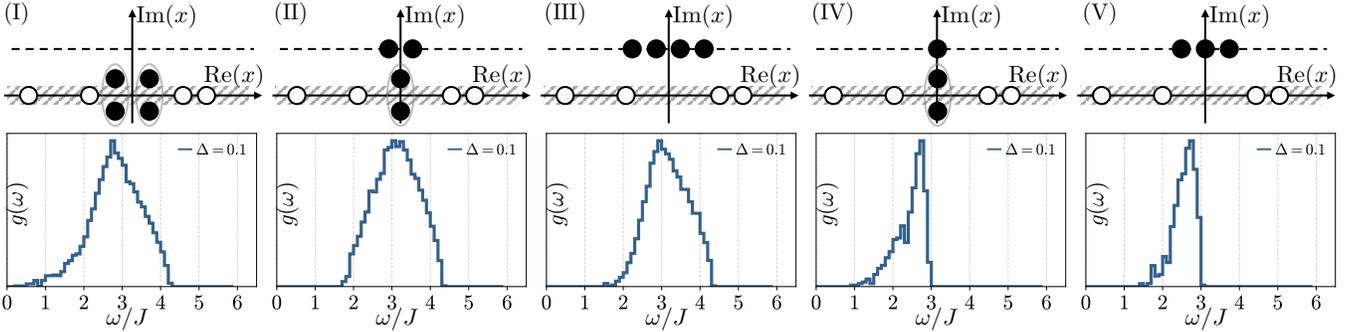

FIG. S5. (I-V) Schematic of the five excitation families discussed in the text. Open circles mark the real BQNs removed from $\{-\frac{M}{2}, \ldots, \frac{M}{2}\}$ or $\{-\frac{M-1}{2}, \ldots, \frac{M-1}{2}\}$; filled circles indicate 2-strings or length-1 complex rapidities. Below each illustration, the density of states $g(\omega)$ is shown for $\Delta = 0.1$. Only the type-I family connects smoothly to the free-fermion limit: its peak matches the free-fermion value, and its spectrum extends continuously to zero energy with no gap as $\omega \to 0$.

As outlined in the previous sections, admissible excitations with complex rapidities or 2-strings have their associated BQNs pinned near zero. As a consequence, the remaining degrees of freedom are carried by the real rapidities, here, by the positions of the holes in the $\nu = +1$ sea (the annihilated real rapidities). Also, changing the total number of real rapidities flips the BQN grid between integer and half-odd-integer labels and reindexes the sea. As a result, admissible excitations relative to the half-filled ground state always involve an *even* number of holes. Therefore, before imposing momentum conservation, the two-hole and four-hole families carry two and four momentum degrees of freedom. Imposing zero total momentum reduces these to one and three. In particular, the one-parameter two-hole family is too restricted to reproduce the excitation types that continue from the free-fermion limit $\Delta = 0$, where the two-hole-two-particle sector has four degrees of freedom (or three at fixed total momentum). We therefore focus on excitation families that retain four momentum variables before imposing the momentum constraint.

The families with four degrees of freedom are:

(I) two 2-strings,

(II) two complex rapidities plus one 2-string,

(III) four complex rapidities,

and their densities of states (DOS) for several $\Delta$ are shown in the main text. There are two additional families:

(IV) one complex rapidity plus one 2-string,

(V) three complex rapidities.

Cases (IV) and (V) leave $M-3$ real rapidities. Since $M-3$ is of opposite parity to $M$, the allowed real rapidities are drawn from the set $\{-\frac{M}{2}, \ldots, \frac{M}{2}\}$ (size $M+1$) rather than $\{-(\frac{M-1}{2}), \ldots, (\frac{M-1}{2})\}$ (size $M$), effectively removing four degrees of freedom. Moreover, for families (IV) and (V), the number of complex rapidities is odd. By Eq. (S90), each complex rapidity contributes an additional momentum $\pi$, so an odd count thereof leaves a residual momentum $\pi$ (mod $2\pi$). Accordingly, Eq. (S86) has to be modified, such that the sum of BQNs for the real rapidities is $M$ (mod $2M$).

Enumerating all consistent BQN assignments, Fig. S5 shows the five families and their DOS $g(\omega)$ at $\Delta = -\Delta_{\mathrm{FM}} = 0.1$. Only family (I) connects continuously to the free-fermion limit. Numerically, we find that the type-V family fails to produce valid solutions at larger $\Delta$ as the Bethe equations do not converge.

--------



\end{document}